\documentclass[nologo,11pt,a4paper]{ETHpaper}     
\usepackage[pdftex]{graphicx}
\usepackage{subfigure}
\usepackage{amsmath}
\usepackage{multirow} 
\usepackage{color, colortbl}  
\usepackage{afterpage}    
\usepackage{url} 
\definecolor{Gray}{rgb}{0.9,0.9,0.9} 
 
\begin{document}   

\title{The Role of Emotions in Contributors Activity:
\\A Case Study of the \textsc{Gentoo} Community }
\titlealternative{The Role of Emotions in Contributors Activity: A Case Study of the \textsc{Gentoo} Community}
\author{David Garcia, Marcelo Serrano Zanetti and Frank Schweitzer}
\authoralternative{David Garcia, Marcelo Serrano Zanetti and Frank Schweitzer}
\address{Chair of Systems Design -- \url{www.sg.ethz.ch} -- ETH Zurich} 
\reference{submitted to the International Conference on Social Computing and Its Applications 2013}
\www{\url{http://www.sg.ethz.ch}} 
\makeframing
\maketitle 

\begin{abstract}

  We  analyze the relation between the 
  emotions and the activity of contributors in the Open Source Software project
\textsc{Gentoo}. Our case study builds on extensive data sets from the project's 
  bug tracking platform \textsc{bugzilla}, to quantify the activity of contributors, and its mail archives, to quantify the emotions of contributors by means of
  sentiment analysis. The \textsc{Gentoo} project is known for a period of centralization within its bug triaging community. This was followed by considerable changes in community organization and performance after the sudden retirement of the central contributor. We analyze how this event correlates with the
  negative emotions, both in bilateral email discussions with  the central
  contributor, and at the level of the
  whole community of contributors.  We then extend our study to consider the activity patterns on 
  \textsc{Gentoo} contributors in general. We find that
  contributors are more likely to become inactive when they express
  strong positive or negative emotions in the bug tracker, or when they deviate from the
  expected value of emotions in the mailing list.  We use these
insights to develop a Bayesian classifier that detects the risk of contributors leaving the project. Our analysis opens new perspectives for measuring online contributor motivation by means of sentiment analysis 
  and for real-time predictions of contributor turnover in Open Source Software projects.

  \end{abstract}

\section{Introduction}
 
Collaboration within an online environment is an everyday challenge
for contributors of Open Source Software (OSS) projects.  They need to interact with other contributors 
to decide about the direction of their project and, equally
important, need to interact with users to learn about their demands.  
Communication within the contributor's community and towards the user's community 
both impact project
reputation and the availability of resources, which are crucial
to further develop the project.  Thus, understanding how people
interact, collaborate and communicate online is an important field of research that
has the potential to improve the performance of OSS projects.
  
The relevance of OSS projects goes beyond research, and reaches wide
industrial applications.  The current technological landscape is
constantly influenced by large OSS projects that generate important
software products.  For example, the \textsc{Apache} server is used in
more than $60\%$ of the
websites\footnote{\url{http://w3techs.com/technologies/details/ws-apache/all/all}},
and \textsc{Firefox} and \textsc{Chrome} have a combined market share
of more than
$50\%$\footnote{\url{http://www.w3counter.com/globalstats.php?year=2013&month=04}}.
These are possible thanks to the efforts of OSS projects, in which
potentially large amounts of contributors can participate by coding,
proposing functionalities, or reporting and triaging bugs.  All
contributors benefit equally from the project, receiving the software
product and its code as a result.  These contributors are free to stop
collaborating at any time; a decision that does not prevent them from
profiting from the project and using its products.  In this sense, an
OSS community is an example of a public goods game
\cite{Andreoni1990}, in which participants have no punishment for
free-riding, and they equally benefit from the common good.  This
poses a paradox, as the game theoretical result of the ``tragedy of
the commons'' \cite{Hardin1968} implies that, when collaborators are
purely rational, the expected outcome of the project is a complete
failure.
      
In proprietary software projects, developers, analysts, and testers
are bound by legal contracts that provide a mechanism to cope with
conflicts and guarantee a certain level of collaboration.  On the
other hand, OSS projects are mostly composed of volunteer
contributors, whose collaboration scheme can be fragile and suffer in
the presence of disagreements or loss of motivation.  For example, the
\textsc{Pidgin} project developed a program for instant messaging
commonly used in \textsc{LINUX}
distributions\footnote{\url{http://www.pidgin.im/}}.  After the
release of a new version, users, developers and other contributors disagreed on a change
related to its user interface, leading to a heated discussion and the
expression of negative
emotions\footnote{\url{https://developer.pidgin.im/ticket/4986}}.  As
a result, the project was divided (i.e. forked) into two different
projects, which is equivalent to a large exodus of contributors. This
example illustrates the impact that the emotional climate of an OSS
community has on its success.  Certain level of positive emotions
seems necessary to sustain the intrinsic motivation of the
collaborators, and strong instances of negative emotions pose a threat
that trigger the turnover of important contributors.

The human factor of OSS projects composes the mechanism that make them
possible, but also poses a threat that endangers their success.
Often, the social component of the projects is analyzed through social
network analysis \cite{crowston2006,Ehrlich2007, He2012,Zanetti2013a},
but the psychological component of OSS interaction has not been
explored so far.  Thanks to the development of tools for sentiment
analysis \cite{thelwall2012sentiment}, we can quantify the emotions of
OSS contributors, looking for relations between their activity and
emotional expression.  Furthermore, given the availability of large
datasets of OSS development forums, this sentiment analysis can be
extended to higher levels of aggregation in which collective emotions
emerge from the interaction of individual contributors.  This poses the
opportunity to empirically analyze the conditions that lead to the
turnover of OSS contributors, and to apply such findings in the
creation of tools to monitor and predict the evolution of OSS projects
\cite{Crowston2012, serrano2012co, valetto2007using}. 
           
In this article, we explore preconditions for contributor turnover,
and their impact on the performance and cohesion of the community.  We
focus on the large \textsc{Gentoo} project, by analyzing two disjoint
datasets spanning about $10$ years of activity and more than 35,000
contributors.  The first dataset contains the records of bug triaging
and processing (i.e. bug tracker) while the second contains messages
exchanged within the developers' mailing list.  To these data, we
apply sentiment analysis, looking for indicators to predict the
turnover of contributors from an OSS community.  Our quantitative
results show that the emotional expression of a contributor is an
indicator for the likelihood of a contributor to remain active in the
project. Finally, we apply these results to formulate a real-time
prediction of contributors leaving the project, providing a tool that
enables timely reactions against undesirable turnover events.
  
\section{Related Work}
  
\subsection{Social Dynamics of Open Source Software}
       
The social organization in open source communities has been addressed
in a number of relevant works.  In \cite{Mockus2002tse}, the focus was
in division of labor.  By analyzing a dataset composed of the
\textsc{Apache} and \textsc{Mozilla} projects, the authors show that
while coding efforts are concentrated on a few \emph{core-developers},
maintenance activities, such as bug report triaging, are performed by
a much larger community.  In \cite{crowston2006}, this core periphery
structure was also studied within a social network analysis framework.
The social network framework was also recently applied by
\cite{He2012} for the study of the behavior of individuals within
communities.  Contributor motivation and its relationship to project
performance is also an important topic, and was considered in a number
of works reviewed in \cite{lerner2002, gacek2004, krogh2006}.
Finally, \cite{cataldo2008} proposes a framework to analyze the
congruence between technical and social organization within a software
project.  In this way, the authors wish to answer the question of
which social organization structure is the best performing given a
particular technological scenario.  Or the analogous, how to structure
a technical architecture in order to fit an established social
organization.

\subsection{Emotions in Social Media}
 
The most common mechanisms for communication in OSS projects are
forums and bug trackers, which are special cases of social media.
This allows the application of sentiment analysis tools
\cite{thelwall2012sentiment}, providing insights to the psychological
experience \cite{Kappas2013} of OSS contributors, rather than just
their social interaction. This approach has been proved useful for the
analysis of collective emotions in forum discussions
\cite{Chmiel2011}, emotional interaction in chatting communities
\cite{Garas2012}, and to test previous hypotheses from psychology in
online data \cite{Garcia2011, Garcia2012b}.  The attention to
sentiment analysis is increasing due to its multiple applications in
finance and marketing. For example, mood measures from social media
have been used to predict the stock market \cite{Bollen2010,
  Deng2011}. Sentiment analysis has also been applied to customer
emotions in Amazon product reviews \cite{Garcia2011e}, and to the
viral spread of information in \textsc{Twitter} \cite{Pfitzner2012}.

Different sentiment analysis tools are available, depending on the
type of analysis and data to process. Supervised methods use training
data to mine emotions and opinions from text \cite{Paltoglou2010}, and
word category frequencies can be used to measure collective mood
\cite{Zhang2012, Bollen2010}. Regarding short and informal text,
lexicon-based classification provides unsupervised methods to extract
sentiment. The state-of-the-art tool in such situation is
\textsc{Sentistrength} \cite{thelwall2012sentiment, Thelwall2013},
which we use in this article. The accuracy of its last version has
been validated with human annotations of a wide variety of online
communities \cite{thelwall2012sentiment}. Among its previous
applications, \textsc{Sentistrength} has been used to analyze emotions
about political topics in \textsc{Youtube} and \textsc{Twitter}
\cite{Garcia2012a, Wu2011}, product reviews \cite{Garcia2011e}, and
\textsc{Yahoo! answers} \cite{Kucuktunc2012}.
    
\subsection{Social Resilience and Contributor Motivation}
 
The question of how groups are formed and disappear has been addressed
for online social networks and scientific communities
\cite{Backstrom2006,Zheleva2009}. This highlights the relevance of
trust networks in social recommender systems \cite{Walter2009}, and
how social movements in \textsc{Twitter} grow and decay through
spreading patterns and complex contagion
\cite{Gonzalez-Bailon2011}.  The departure of
individual users, commonly denoted as churn, has also been analyzed
for the online communities like \textsc{Yahoo! answers}
\cite{Dror2012}, and other social networks \cite{Wu2013}.
Furthermore, previous analysis provide insights on the decision of
users to leave P2P networks \cite{Herrera2007}, discussion boards
\cite{Karnstedt2010}, and online videogames \cite{Kawale2009}.  These
previous works focused on the relation between social indicators, like
amount of contacts, with the likelihood of users to leave an online
community. While useful as a first approximation, these analyzes did
not take into account emotional expression and interaction, which are
related in the psychology literature to motivation and social
interaction \cite{hall1961psychology, lerner2002, krogh2006}.
    
The microscopic dynamics that drives the decisions of users to leave a
community create the macroscopic effect of social resilience
\cite{Garcia2013a}, or how strong is the community when facing
disrupting periods. Such disrupting events have been characterized by
text analysis on \textsc{Facebook} \cite{Cvijikj2011}, but their
influence on the survival of an online community at large cannot be
simply mapped to its social network \cite{Garcia2013a}. The intrinsic
motivation of the users and their individual decisions are key factors
for the collective dynamics of the community. As an example, external
incentives do not guarantee more efficient viral marketing campaigns
\cite{Michalski2012}. On the other hand, information spread can be
motivated by emotional content \cite{Pfitzner2012}, leading to higher
levels of user activity and interaction when emotions are involved.

\section{\textsc{Gentoo} Datasets}
 
\subsection{Bug reports} 
  
The \textsc{Gentoo} project adopts the \textsc{Bugzilla} as its bug
tracking system \cite{serrano2005}.  It is composed of an online
database\footnote{\url{https://bugs.gentoo.org/}} where each entry is
organized around the notion of a \emph{bug report}.  A bug report
status will change as its processing progresses towards a solution
(e.g. \emph{pending}, \emph{reproduced}, \emph{closed}, etc).  In
general, the modification of a bug report field (e.g. status) allows
its author to leave simple text comments.  Using the \textsc{Bugzilla}
API, we collect the time series of comments, along with the unique
\emph{username} of its author and the unique \emph{id} of the
respective bug report.  In Table \ref{tab:Statistics}, we summarize
the main statistics of this dataset.  To each of those comments, we
apply the \textsc{Sentistrength} tool in order to quantify its
positive and negative valence.
   
\begin{table}[htpb] 
\centering
\caption{Basic statistics of the datasets used for this study. We cover activity within \textsc{Gentoo}'s bug tracker and within the \textsc{gentoo-dev} mailing list\label{tab:Statistics}.} 
\begin{tabular}{l|r|r} 
     Statistics & \textsc{Gentoo} \textsc{Bugzilla} & \textsc{gentoo-dev}\\
     \hline
     \rowcolor{Gray}                                         & 01/04/2002& 04/01/2001\\
     \rowcolor{Gray}    \multirow{-2}{*}{Observation period}& to 04/26/2012 & to 29/06/2012 \\
       Messages                            & 661,783  & 81,328 \\
  
     \rowcolor{Gray} Discussions                         & 140,216  & 14,070 \\  
                        Contributors                        & 36,555 & 4,664 \\
   \end{tabular}
\end{table} 

\subsection{Developer mailing list} 

While triaging and processing bug reports, contributors may rely on
information exchange through mailing lists.  In the case of
\textsc{Gentoo}, this is mainly done via the \emph{gentoo-dev}
list\footnote{\url{http://archives.gentoo.org/gentoo-dev/}}, which is
the list subscribed by core-developers and code maintainers.  Thus if
contributors processing bug reports want to call the attention of a
serious maintainer, that is the best place to start.  Messages sent to
this mailing list are stored in a database and can be retrieved at any
time from their archive version, which is accessible via a HTML
interface.  Using this channel, we extract the time series of email
messages sent to the \emph{gentoo-dev} list, along with the unique
\emph{userid} of its author and message subject, which repeats for all
messages sent to the same thread.  Again, the textual content of each
message is analyzed with \textsc{Sentistrength} yielding positive and
negative valence scores (i.e. excluding content commented out with
character ``$>$'' at the start of each new line).

\subsection{Sentiment analysis}

We process all comments and messages in the bug reports and the
developer's mailing list using \textsc{Sentistrength} \cite{thelwall2012sentiment}.
\textsc{Sentistrength} is the state-of-the-art tool for lexicon-based analysis
of social media messages, in particular for informal communication.
It has been validated on test datasets including \textsc{Digg} and
other fora on specialized topics, which are similar communication
media as the \textsc{Gentoo} bug tracker and mailing list.  When classifying
the polarity of forum messages, \textsc{Sentistrength} has an accuracy above
$88\%$ for \textsc{Digg}, and above $90\%$ in other fora
\cite{thelwall2012sentiment}. It has high correlation values with human raters
on these communities, providing sentiment scores that would be
indistinguishable from a human rater, and providing not only an
accurate, but also a \emph{valid} estimation of the sentiment. For
these reasons, previous works have applied it to \textsc{Yahoo! answers} \cite{Kucuktunc2012}, \textsc{Twitter} messages
\cite{Pfitzner2012}, and chatroom communication \cite{Garas2012}.
   
\textsc{Sentistrength} uses a lexicon of emotional-bearing terms combined with
the detection of negations, amplifiers and diminishers.  Its output is
composed of two values, a measure of positive sentiment $p \in
[+1,+5]$, and a measure of negative sentiment $n \in [-1,-5]$.
Following the rationale of \cite{thelwall2012sentiment}, we can aggregate these
two values to a measure of polarity. A message $m$ is classified as
positive ($s = +1$) if $p+n > 0$, negative ($s = -1$) if $p+n <
0$, or neutral ($s = 0$) if $p=n$ and both have an absolute value
lower than $4$.  Comments with high and equal positive and negative
sentiment do not map to this unidimensional
simplification. Nevertheless, this approximation is valid in our data
analysis, as only $265$ messages were detected as
in the cases of $[+4,-4]$ or $[+5,-5]$. These messages compose $0.032\%$ of
the total, and we discard them from our analysis.

\begin{table}[htpb] 
\centering 
\caption{Message ratio per polarity within \textsc{Gentoo}'s bug tracker and within the \textsc{gentoo-dev} mailing list\label{tab:StatisticsPol}.} 
\begin{tabular}{c|c|c}
     Polarity & \textsc{Gentoo} \textsc{Bugzilla} & \textsc{gentoo-dev}\\
     \hline
     \rowcolor{Gray}   positive                   &0.28 &0.28  \\
                        neutral                   &0.56 &0.49  \\
     \rowcolor{Gray}    negative                  &0.16 &0.23  \\
\end{tabular}
\end{table} 

\section{The departure of a central contributor} 
       
In this section we quantify and discuss the role of emotions in a case
study focused on the \textsc{Gentoo}-\textsc{Linux} project.  The
\textsc{Gentoo} project is of particular interest due to a well
documented centralization event followed by significant changes in community performance \cite{Zanetti2013b}.  In that work, we focus
on the evolution of social organization within \textsc{Gentoo}'s bug
triaging community. Using a quantitative methodology based on social
network analysis \cite{Zanetti2012, Zanetti2013a}, we show that we can
monitor drastic changes in social organization which are usually
associated with increased risks.  More specifically, the bug triaging
community of \textsc{Gentoo} came to rely on a single person
(i.e. named \emph{Alice}) to help them in processing bug reports.
This is in accordance with previous findings relating centrality to
preference in collaboration \cite{He2012}.  Based on \emph{Alice}'s
activity, we divide the timespan of our dataset into three
observation periods $P1$, $P2$, $P3$.  In period $P1$, between January
2002 and October 27, 2004, \emph{Alice} was not yet active and the
community was growing.  During the second period $P2$ starting on
October 28 2004, \emph{Alice} gradually became the most central
contributor.  She unexpectedly left the community after her last
contribution on March 29 2008, which marks the start of the third
period $P3$ in which \emph{Alice} was not active anymore.  In the next
we discuss \emph{Alice}'s impact on community performance and the
possible effects of emotions on her motivation to leave the project.
   
\subsection{Effect in performance}
         
During $P2$, \emph{Alice} concentrated most of the work related to bug
triaging on herself, and as a result, the time to first reply and
finally solve open bug reports were minimized.  These are important
metrics that correlate to the likelihood of a bug reporter in becoming
a \emph{long time contributor} to the project \cite{mockus2012}.  The
main issue about \emph{Alice}'s impact was that -- due to personal
conflicts, and dissatisfaction with the social environment of the
project as a whole -- she left the community suddenly.  As we show in
\cite{Zanetti2013b}, after \emph{Alice}'s retirement (i.e. period
$P3$) the community never managed to achieve the same levels of
performance.  Thus, besides monitoring changes in community social
organization and its implied risks, we wish for quantitative measures
that could give an early indication to individual loss of
motivation or activity.
      
\begin{figure}[t]     
\centering
\includegraphics[width=\textwidth]{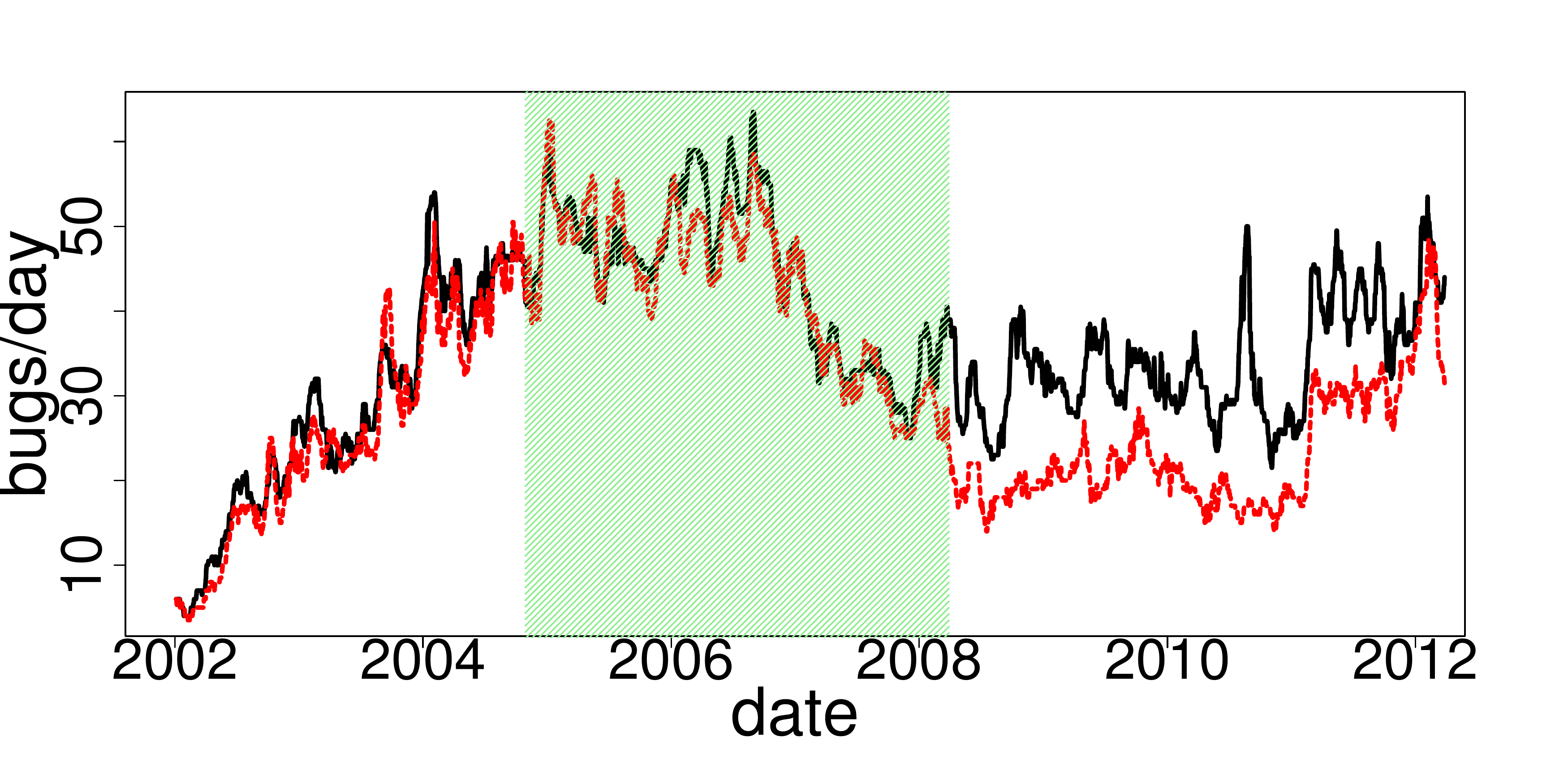} 

\includegraphics[width=\textwidth]{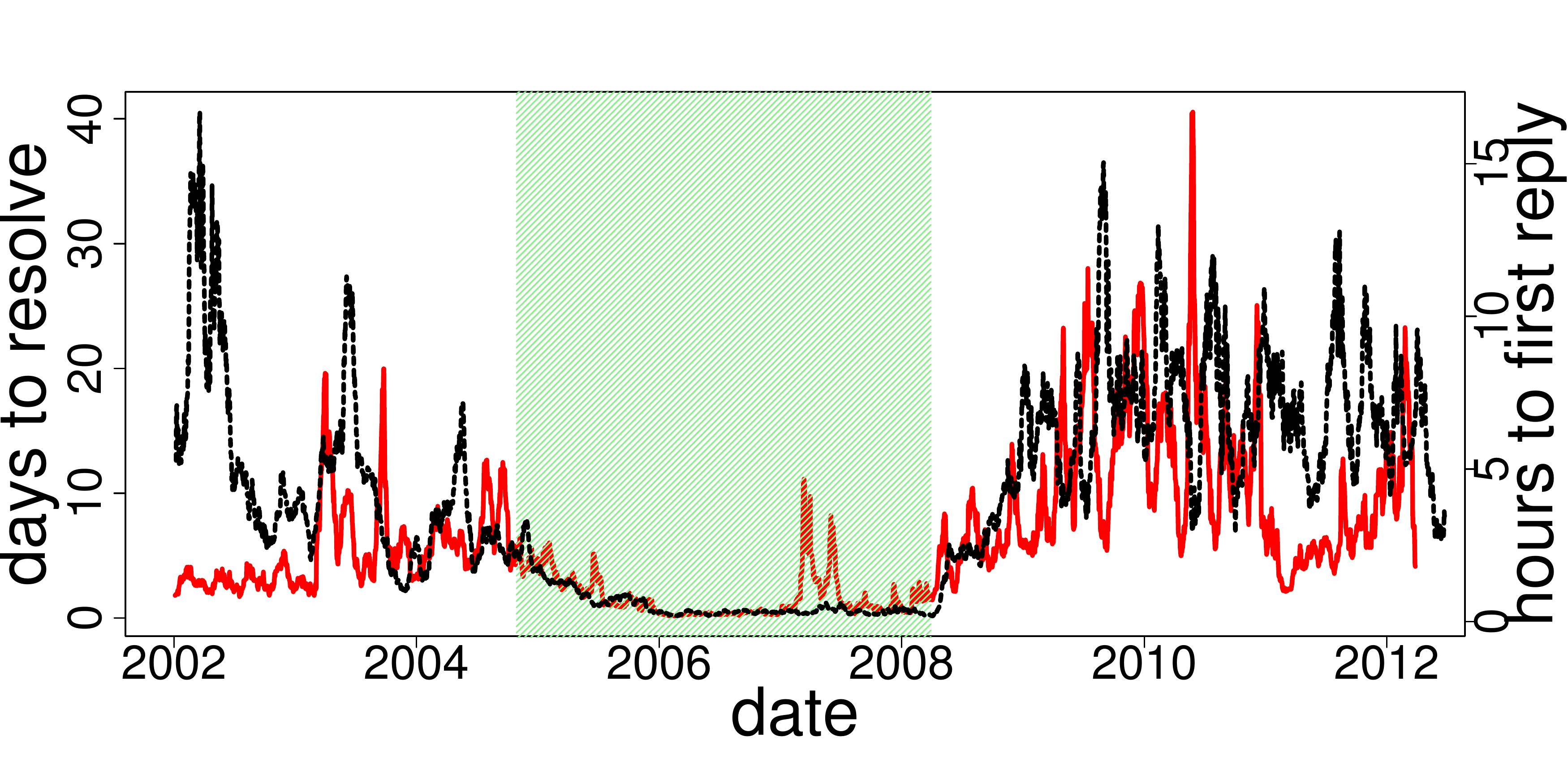}
\caption{Time series of performance metrics. On top, the median values of bugs reported (black curve) and solved (red curve) per day. On bottom, the median time to solve (red curve) and give the first reply (black curve) to a bug report.
The green interval highlights $P2$, the period when \emph{Alice} was active.
}
\label{fig:performance}     
\end{figure} 
   
\subsection{Changes in collective emotions}
 
\begin{figure}[!ht]        
\centering   
  \includegraphics[width=1\textwidth]{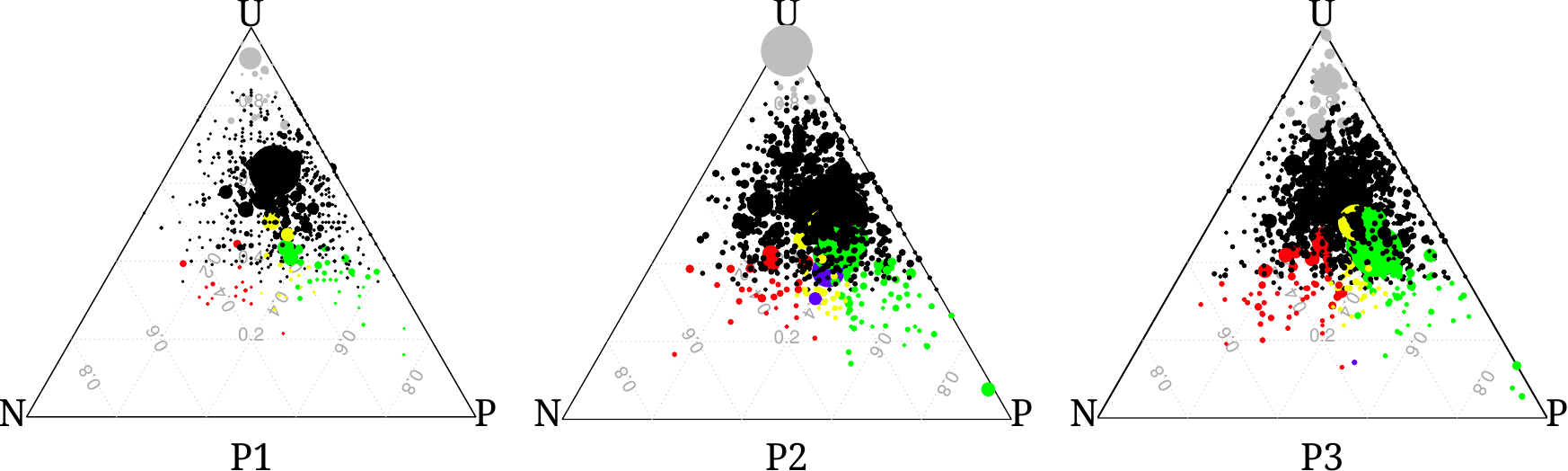}

  \vspace{0.1cm}

  \includegraphics[width=1\textwidth]{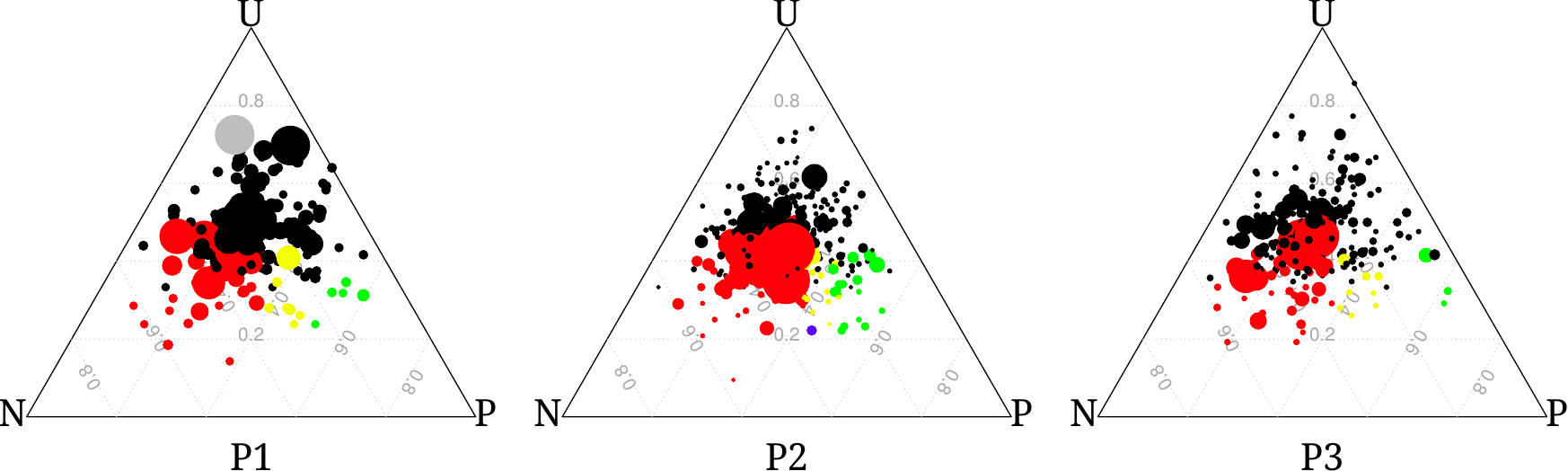}
  \caption{Triangular representation of the emotions in \textsc{Gentoo} discussions in bug tracker (top) and the developer mailing list (bottom). 
Points represent discussions with at least $20$ messages, with a size proportional to the amount of messages in the discussion, at a distance to the triangle vertices proportional to the ratios of positive, negative and neutral messages. 
Points are colored according to the classification of the discussion (i.e. black for \emph{neutral}, green for \emph{positive}, red for \emph{negative}, gray for \emph{underemotional}, blue for \emph{bipolar} and yellow for \emph{undetermined})\label{fig:triangles}.}
\end{figure}

To measure the collective emotions in the discussions associated with comments to a bug report or e-mails to a thread in the mailing list, we aggregate the emotional values of the messages in the discussion. 
In this way, for the set of messages in a discussion $M_d$, we calculate the ratios of positive $P_d = \frac{\sum_{m \in M_d} s_m = 1}{|M_d|}$ , negative $N_d=\frac{\sum_{m \in M_s} s_m = -1}{|M_d|}$, and neutral $U_d =\frac{\sum_{m \in M_s} s_m = 0}{|M_s|}$ messages.  
These measurements map the discussions to a simplex on the plane \cite{Garcia2012a}, where each discussion has a distance to the vertices of a triangle proportional to $P_d$, $N_d$, and $U_d$.  
Figure \ref{fig:triangles} shows this representation separately for bug tracker and the mailing list, where each discussion $d$ is a point of size proportional to $|M_d|$. 
 
The ratio of the overall emotional expression in each medium (i.e. bug tracker or mailing list), $\bar{P}, \bar{N}, \bar{U}$, allow us to compare the emotions of a discussion with this ground state of the \textsc{Gentoo} community. 
We perform a set of nonparametric statistical tests to classify each discussion, consisting on three $\chi ^2$ tests at the $95\%$ confidence interval:
   
\begin{enumerate}
\item Test of $U_d \simeq \bar{U}$: if this hypothesis cannot be rejected, the discussion is not considered to include collective emotions, and it is classified as \emph{neutral}. 
If the $U_d > \bar{U}$ hypothesis is supported, we classify the discussion as \emph{underemotional}. 
Examples of this kind of discussions are exchanges of computer code or error logs, which serve a technical purpose but do not compose emotional interaction. 
If the hypothesis $U_d < \bar{U}$, is supported, the discussion contained collective emotions, and the next two tests are evaluated to classify the
  emotions in this discussion.
\item Test of $P_d \simeq \bar{P}$: if the null hypothesis can be
  rejected and the data supports $P_d > \bar{P}$, we classify the
  discussion as \emph{positive}.
\item Test of $N_d \simeq \bar{N}$: in the same way as the previous
  point, if the data supports $N_d > \bar{N}$, we classify the
  discussion as \emph{negative}.
\end{enumerate}
    
The above set of tests allows us to detect discussions that simultaneously contain positive and negative emotions, which will pass the second and third test. 
We classify these discussions as \emph{bipolar}, representing collective emotions in which the authors of messages are polarized in different emotional states \cite{Schweitzer2010}. 
Additionally, a discussion might pass the first test, but not the second nor the third. 
These discussions contain more emotional content than the average of the community, but there is not enough data to classify the polarity of the emotions expressed in it. 
We classify these as \emph{undetermined}.

Our statistical analysis highlights the presence of strong positive discussions in the bug reports, represented by points close to the lower left corner of the triangle. 
In these discussions, positive collective emotions are usually created as the result of fixing a software issue. 
The bug report system also shows some instances of underemotional discussions, represented by gray points close to the upper corner of the triangle. 
These threads are large exchanges of error logs and program outputs, and do not constitute a significant source of emotional interaction.

The lower row of Figure \ref{fig:triangles} shows the collective emotions in the discussions of the developer's mailing list (i.e. \textsc{gentoo-dev}). 
The structure of the emotions in these discussions is significantly different when compared against their bug reports counterpart: there are very few instances of positive discussions, and there are large discussions that elicited negative collective emotions. 
  
  
These differences are possible due to the fact that 
these two communication channels (i.g. bug tracker and mailing list) shows very different styles of emotional interaction.
In the bug tracker, positive emotions prevail.
Users, developers and other contributors need to interact focusing on solving existing software issues. 
Thus, bug reports must be written as clear as possible. 
Moreover, contributors might need to write back to bug reporters in order to gather further information.
This needs to be done in a smooth way that will lead to the identification of the locus of software issue as fast as possible.
On the mailing list the situation can be quite different.
Specially in the case of the developers' private list, large instances of negative emotions can be observed. 
Likely, this is due to disagreements in collaboration processes and on competing agendas specifying how work and software should be organized. 
  
The representation of collective emotions in Figure \ref{fig:triangles} is useful to detect discussions that could trigger the decision of contributors to stop contributing to the OSS community. 
When comparing the three intervals depending on \emph{Alice}'s presence, it is difficult to find differences in this representation. 
Periods $P2$ and $P3$ seem slightly more emotional, with some instances of bipolar discussions. 
For the case of the mailing list, negative emotions appear to be more salient in $P2$ and $P3$, but these observations require a quantitative validation. 
For that reason, we compute the time series of emotions in messages, using a moving average with $T=30$ days range. Thus, $M_T$ represents all messages found within such a time window.
This allows us to calculate the respective mean positivity $p(t) = \sum_{m \in M_{T}} p_m / |M_{T}|$, mean negativity $n(t) = -\sum_{m \in M_{T}} n_m / |M_{T}| $, and mean polarity $s(t) = \sum_{m in |M_{T}|} s_m / |M_{T}|$ of messages.
 
\begin{figure}[!ht] 
\centering
  \includegraphics[width=1\textwidth]{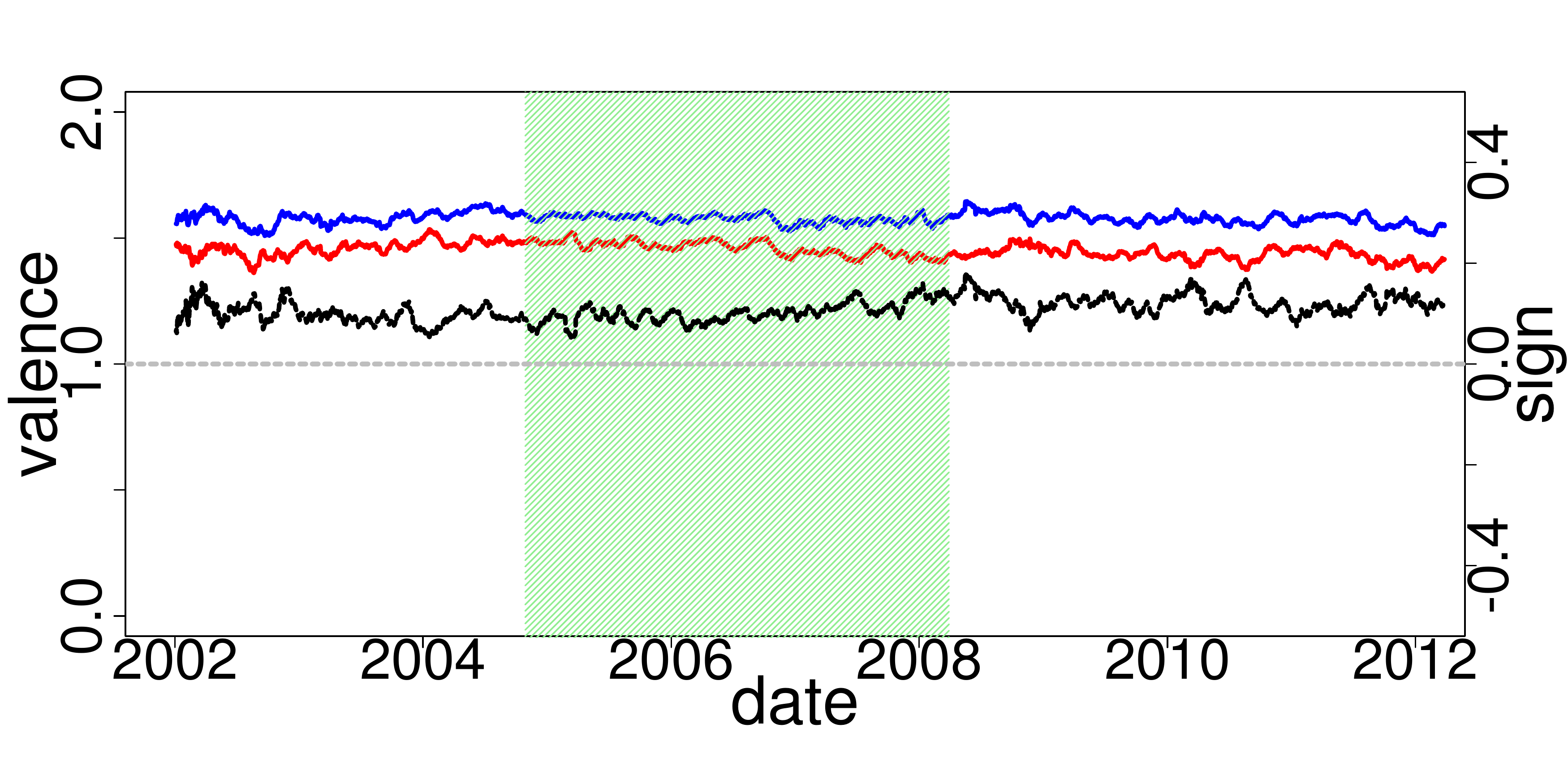} 
  \includegraphics[width=1\textwidth]{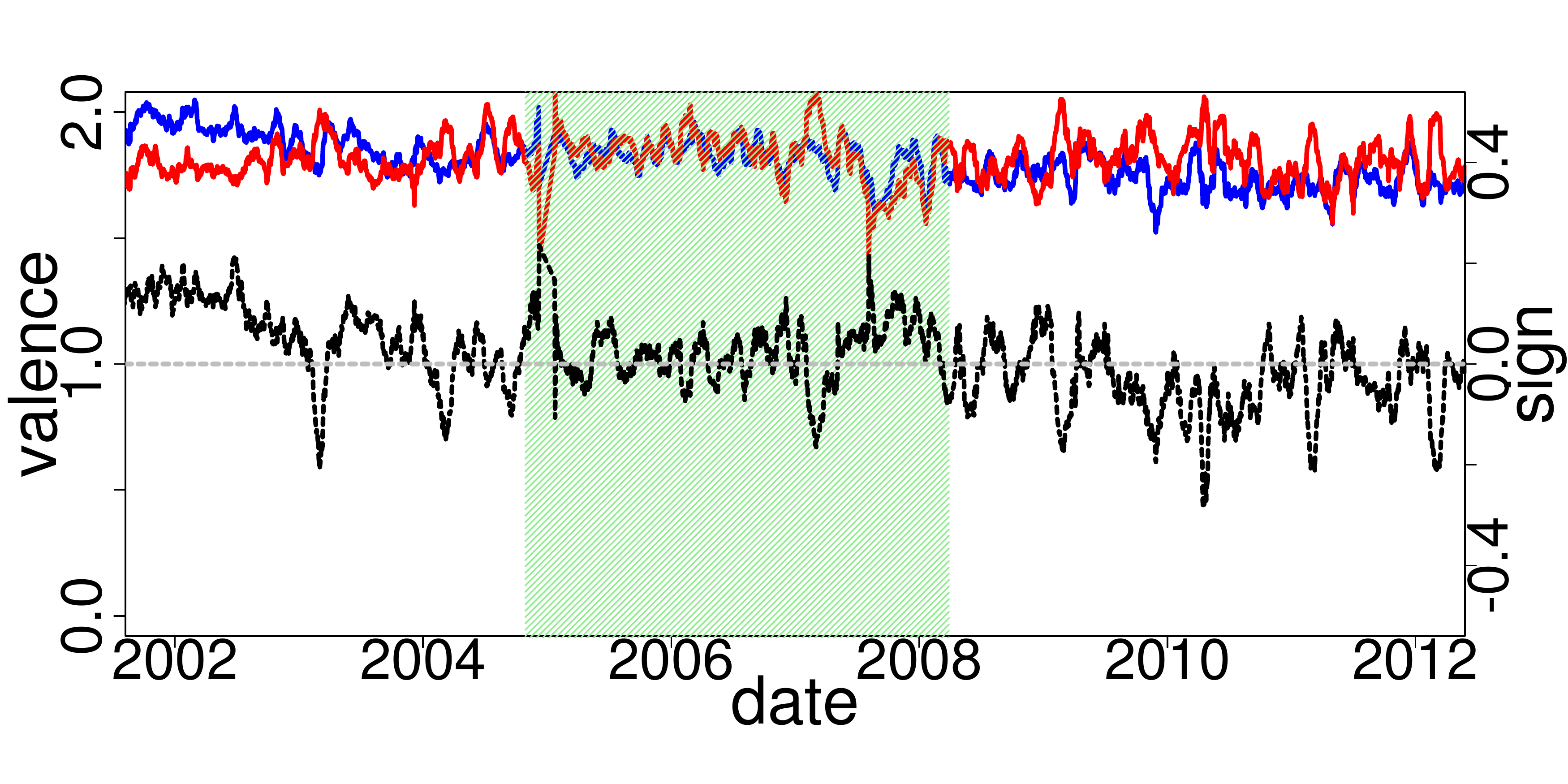}
\caption{Moving average applied to the time series of emotional expression within \textsc{Gentoo} \textsc{Bugzilla} (top) and \textsc{gentoo-dev} (bottom). 
The red curve shows the daily mean negativity in the messages, the blue curve shows the mean positivity, and the black curve represents the mean polarity of all the messages. 
The green interval highlights $P2$, the period when \emph{Alice} was active.\label{fig:EmoTS}} 
\end{figure} 
 
Figure \ref{fig:EmoTS} shows these time series for the bug reports and for the developer's mailing list, divided into the three periods mentioned above. 
It can be noticed that there is no clear effect of the presence or absence of \emph{Alice} in the bug tracker, but the developer's mailing list seems to change. 
Before \emph{Alice}'s presence, the mean polarity used to have positive values, and during her activity this was close to $0$. 
After her departure, there seems to be a period of stronger negativity.
We statistically tested this observations, performing $\chi^2$ test on the values of $s(t)$ across periods.
           
The results of these tests are reported in Table \ref{tab:testprop}, supporting our observation that -- in the mailing list -- $P3$ had more negative emotions than the
periods $P1$ and $P2$ together.    
During that period, the community went through a complete reorganization, catalyzed by the creation of the \emph{bug wranglers} project\footnote{\url{http://www.gentoo.org/proj/en/qa/bug-wranglers/}}.
This was an initiative specially meant to cope with \emph{Alice}'s sudden retirement.    
Thus, the negativity observed during $P3$ is likely to be due to the community struggle in restructuring its procedures.
What about \emph{Alice}'s presence during $P2$?
Did \emph{Alice} experience different sentiment expression within the bug tracker and mailing list?
We separate the discussions in which \emph{Alice} took part, from the remaining taking place within that period, and again calculated the different proportions of polarities.
We show in Table \ref{tab:testprop} that the discussions in the mailing list that contained \emph{Alice}'s messages were indeed more negative and less positive than the discussions not containing her messages, while the proportions of neutral polarity were roughly the same. 
Now focusing on the bug tracker, we observed that the proportion of negative polarity  were roughly the same in \emph{Alice}'s discussions when compared to the remaining discussions.
Moreover, \emph{Alice}'s discussions were more neutral and less positive. 
             
\begin{table}[hptb] 
   \centering         
\caption{Test for statistical significance of differences in proportion of polarities. 
$N$ represents the proportion of negative messages, $P$ for positive ones and $U$ for neutral ones. 
The \emph{null hypothesis} is always $Prop_1$=$Prop_2$. 
The subscripts \emph{P1-P2} and \emph{P3} corresponds to the analysis per period, while \emph{without Alice} and \emph{with Alice} to the analysis per thread.\label{tab:testprop}}  
\begin{tabular}{c}
\textsc{Gentoo} \textsc{Bugzilla}\\ 
\end{tabular} 

\begin{tabular}{c|c|c}
\hline   
p-value of null hypothesis  & alternative hypothesis &  estimate\\
\hline 
 \rowcolor{Gray}             $1.04e-033$ & $N_{\textrm{P1-P2}}$ $>$ $N_{\textrm{P3}}$ &$0.011$\\
     $2.12e-003$ & $U_{\textrm{P1-P2}}$ $>$ $U_{\textrm{P3}}$&$0.003$\\
  \rowcolor{Gray}            $1.29e-040$ & $P_{\textrm{P1-P2}}$ $<$ $P_{\textrm{P3}}$&$0.014$\\
     \hline
     $2.00e-002$ & $N_{\textrm{without Alice}}$ $<>$ $N_{\textrm{with Alice}}$&$0.003$\\
 \rowcolor{Gray}             $2.06e-130$ & $U_{\textrm{without Alice}}$ $<$ $U_{\textrm{with Alice}}$&$0.045$\\
     $6.62e-188$ & $P_{\textrm{without Alice}}$ $>$ $P_{\textrm{with Alice}}$&$0.049$\\
\hline  
    \end{tabular}
 
\begin{tabular}{c}
\textsc{gentoo-dev}\\ 
\end{tabular} 

\begin{tabular}{c|c|c}
\hline   
p-value of null hypothesis  & alternative hypothesis&estimate \\
\hline 
\rowcolor{Gray}    $1.49e-021$ & $N_{\textrm{P1-P2}}$ $<$ $N_{\textrm{P3}}$&$0.033$\\
                    $5.08e-006$ & $U_{\textrm{P1-P2}}$ $<$ $U_{\textrm{P3}}$&$0.017$\\
                    \rowcolor{Gray}     $7.61e-046$ & $P_{\textrm{P1-P2}}$ $>$ $P_{\textrm{P3}}$&$0.050$\\
  \hline
   $1.61e-026$ & $N_{\textrm{without Alice}}$ $<$ $N_{\textrm{with Alice}}$&$0.066$ \\
  \rowcolor{Gray}                  $5.50e-001$     & $U_{\textrm{without Alice}}$ $<>$ $U_{\textrm{with Alice}}$&$0.004$\\
   $8.56e-023$ & $P_{\textrm{without Alice}}$ $>$ $P_{\textrm{with Alice}}$& $0.001$\\
\hline
\end{tabular} 
      \end{table}

\section{Emotions and inactivity}

The above analysis of the departure of \emph{Alice} serves as an
example of the interplay between emotions and activity in the \textsc{Gentoo}
community. In this section, we extend that analysis to contributors in
general, exploring the role of emotions in their activity patterns. We
continue by developing a method to predict long periods where an
individual contributor is inactive.

\subsection{Activity modes} 

For the case of \emph{Alice}, determining when she became inactive is
a trivial task, as she had no activity after a certain date. This is
not necessarily the case for contributors in general, who might be
inactive for a long period and then become active again. In general,
contributors do not have a standard mechanism to inform the rest of
the community if they are active or not, and the only way to detect
their inactivity is when they do not produce messages for a period of
time. To detect if a contributor became inactive, we use the theory of
interevent time distributions \cite{Barabasi2005, Wu2010}, which
divides human communication in two modes: A bursty, \emph{correlated}
mode in which the time between the actions of a human is very short;
and an \emph{uncorrelated} mode that corresponds to the long times
between bursts of activity. The \emph{correlated} mode can be detected
when the interevent times of a human follow a power-law distribution
\cite{Barabasi2005, Garas2012}, which emerges when humans reply to
each other. The \emph{uncorrelated} mode can be detected as an
exponentially decaying regime, which can be explained as the result of
a Poisson process of decoupled actions that start activity bursts
\cite{Wu2010}.

\begin{figure}[!ht] 
\centering
\centerline{ 
  \includegraphics[width=0.5\textwidth]{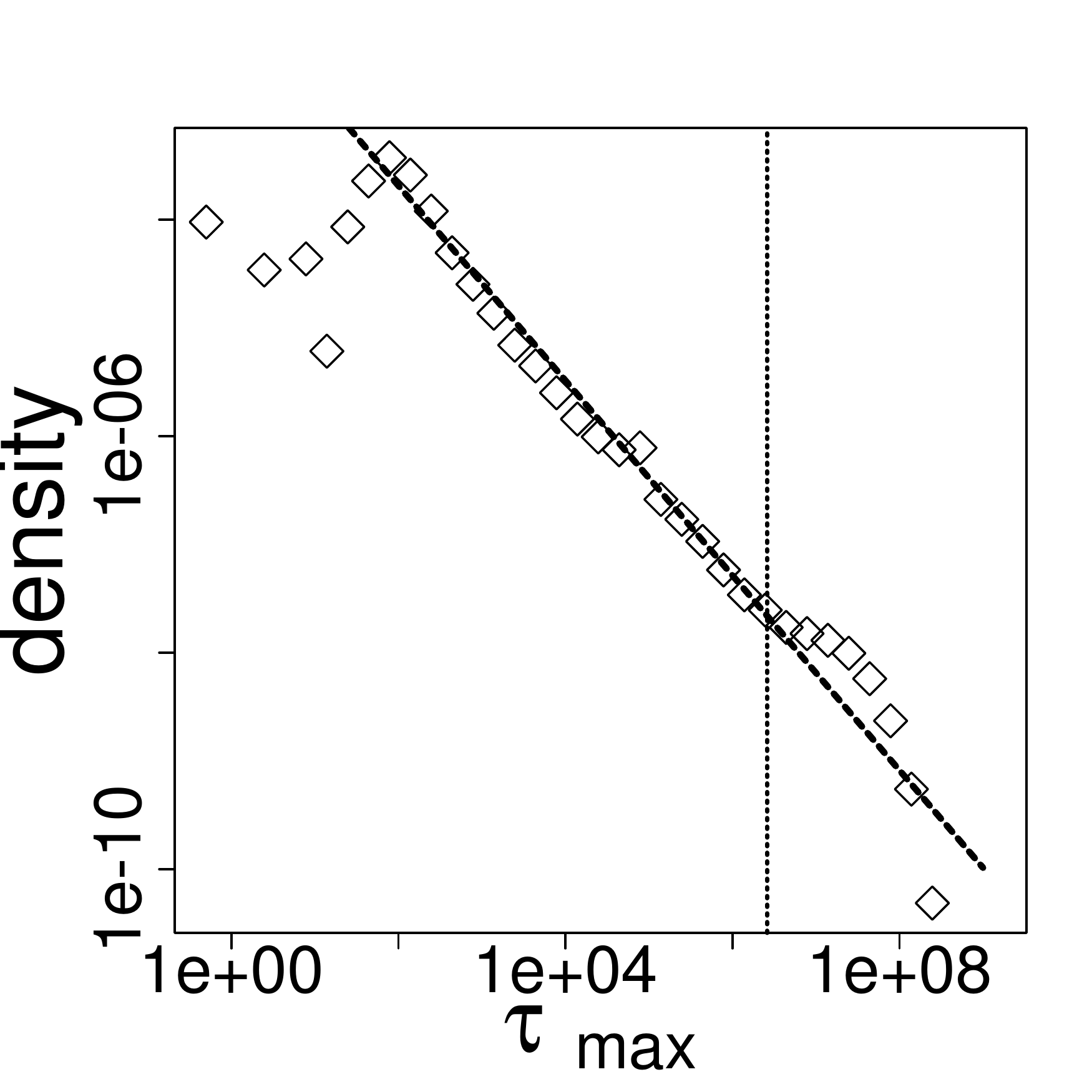}
\hfill  \includegraphics[width=0.5\textwidth]{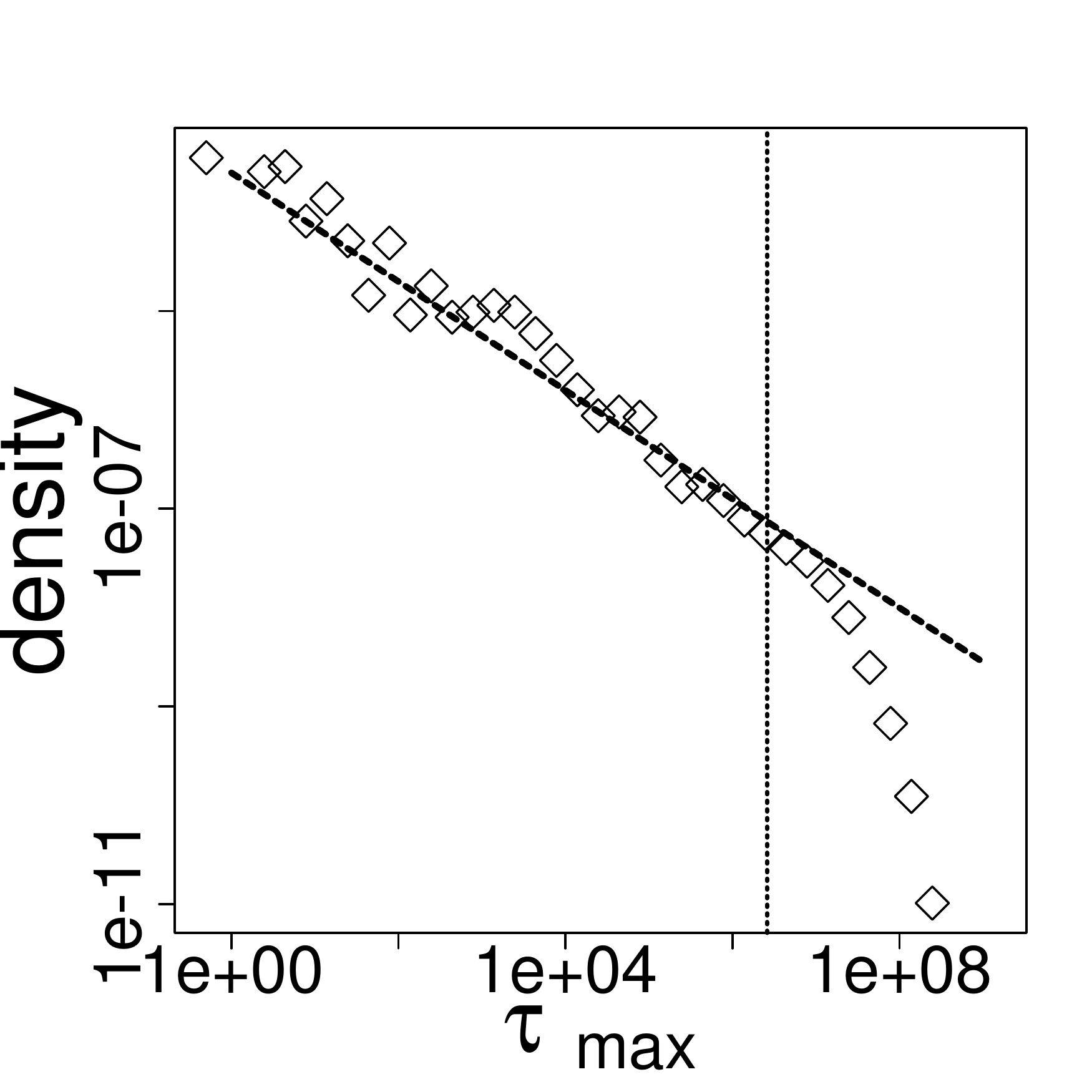}}
\caption{Distributions of the maximum interevent times per contributor
  in the bug tracker (left) and the mailing list (right).  The
  distributions are plotted in a log-log scale with exponentially
  increasing bins, and power-law fits to the head of the
  distributions. Dashed vertical lines show the mark at $\tau_{max}=30$ days,
  where there is a regime change. \label{fig:interEvents}}
\end{figure}

For both datasets, we measured the interevent times $\tau$ between the
messages of each contributor, and characterized the maximum inactivity
period of each one through the maximum interevent time
$\tau_{max}$. Figure \ref{fig:interEvents} shows the distribution of
$\tau_{max}$ for the mailing list and the bug tracker, with power-law
fits to the head of the distributions.  It can be noticed that these
power-law regimes are not valid after a certain value, where both
distributions show a tail that decreases much faster than a
power-law. This shows the division between the two modes of activity
mentioned above: the head of the distributions correspond to the
\emph{correlated} mode of active contributors, while their tails
represent the \emph{uncorrelated} time intervals when contributors are
inactive.  We found that the point between both modes is approximately
$\tau=30$ days. This indicates that when a contributor does not create any
new message for a month, its behavior is uncoupled from the rest and
it can be considered as inactive.

\subsection{Contributor emotions}

To produce the dataset that classifies contributor activity, we do
the following: We collect all messages written by each contributor
$u$, sorting messages by date.  Then we iterate over the messages,
starting from the earliest.  If the contributor only posted a single
message, we discard this \emph{one time contributor} from our analysis.  If
the contributor posted more than one message, for each message $m^u_t$
posted at time $t$, we measure the time interval $\tau$ between it and his
next message $m^u_{t+\tau}$.  If this time interval is shorter than
$30$ days, we label the interval $I^u_t$ as \emph{ACT}, meaning the
contributor is active.  Otherwise, we label the interval as
\emph{INA}, meaning that the contributor started a period of
inactivity according to the theory explained above. For this
analysis, we discard the last message posted by each contributor.

For each interval of contributor $u$, we compute the mean positivity
score $P_u$ and mean negativity score $N_u$ of the messages of the
contributor in the $5$ days preceding the interval. This way, each data
point is an interval between messages of the same contributor, with
real-time measurements of the emotions expressed by that contributor
in the days before the interval takes place. Our aim is to provide a
predictor that identifies when a contributor is going to become
inactive, as a tool that can warn community managers about the risk of
losing contributors. This is not a simple task, as the ratios of each
type of interval are very unevenly distributed. The prior probability
of an interval being labeled as \emph{INA} is $0.088$ in the bug
tracker, and $0.075$ in the mailing list.

\begin{figure}[!ht] 
\centering
\centerline{
  \includegraphics[width=0.5\textwidth]{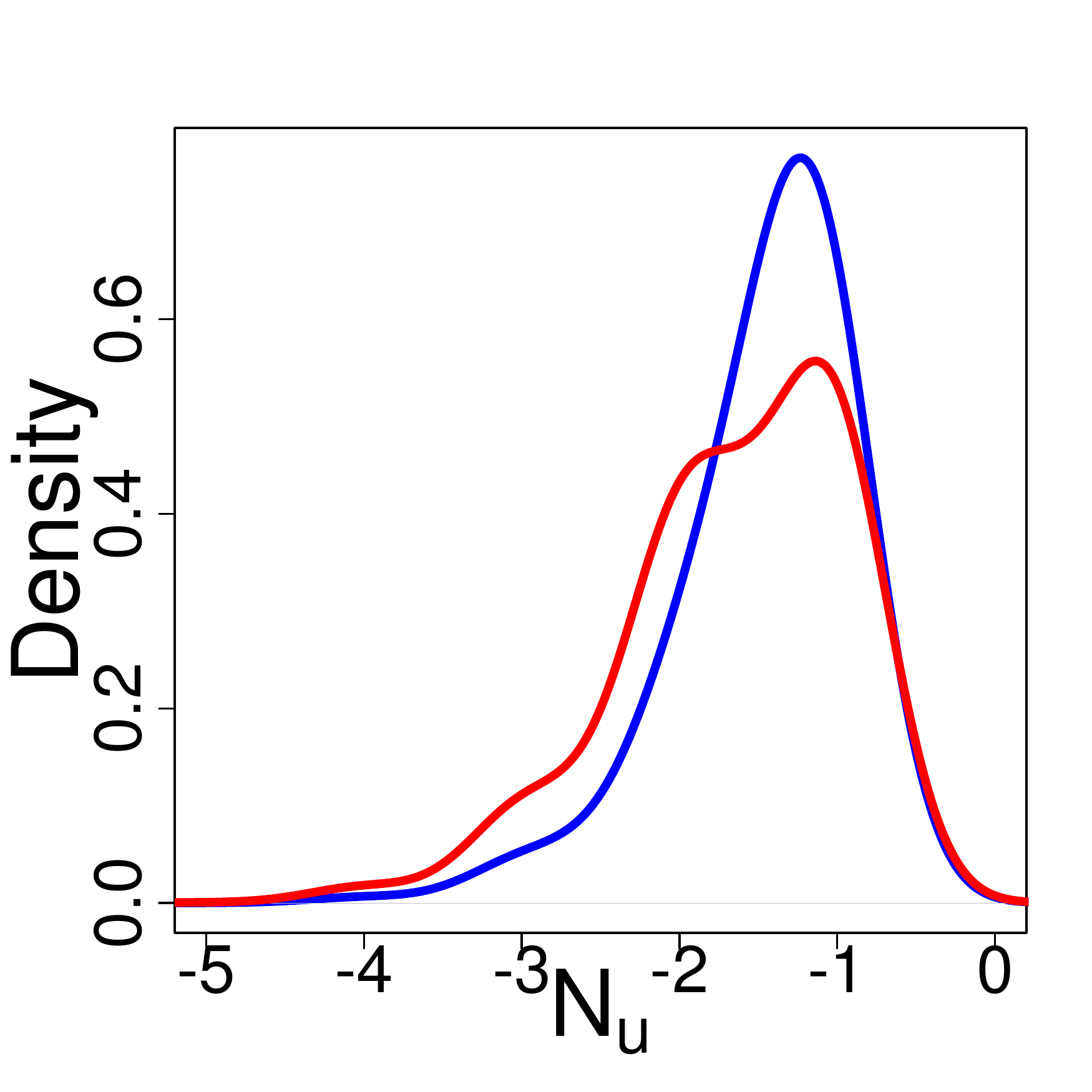}
\hfill  \includegraphics[width=0.5\textwidth]{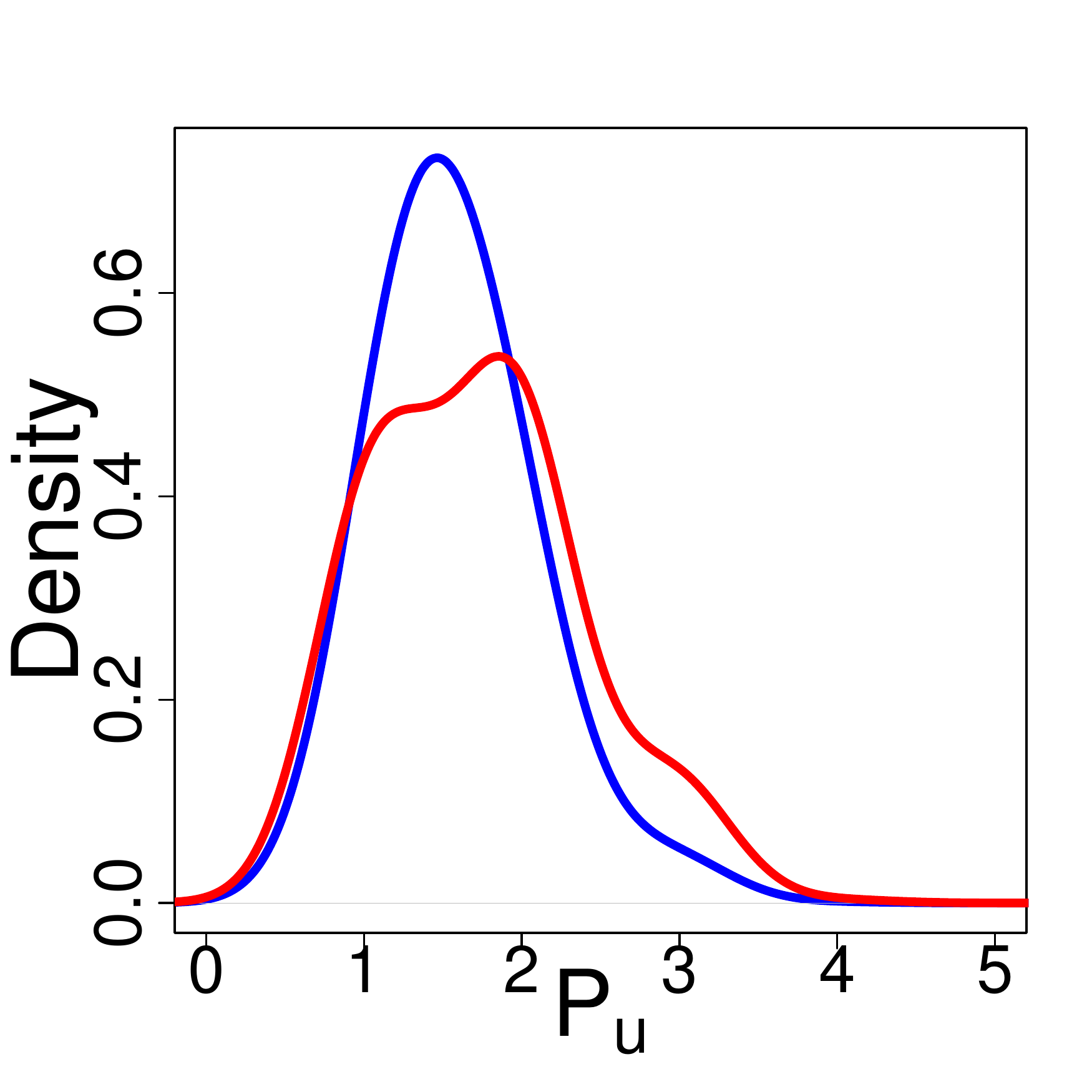}}
\centerline{
  \includegraphics[width=0.5\textwidth]{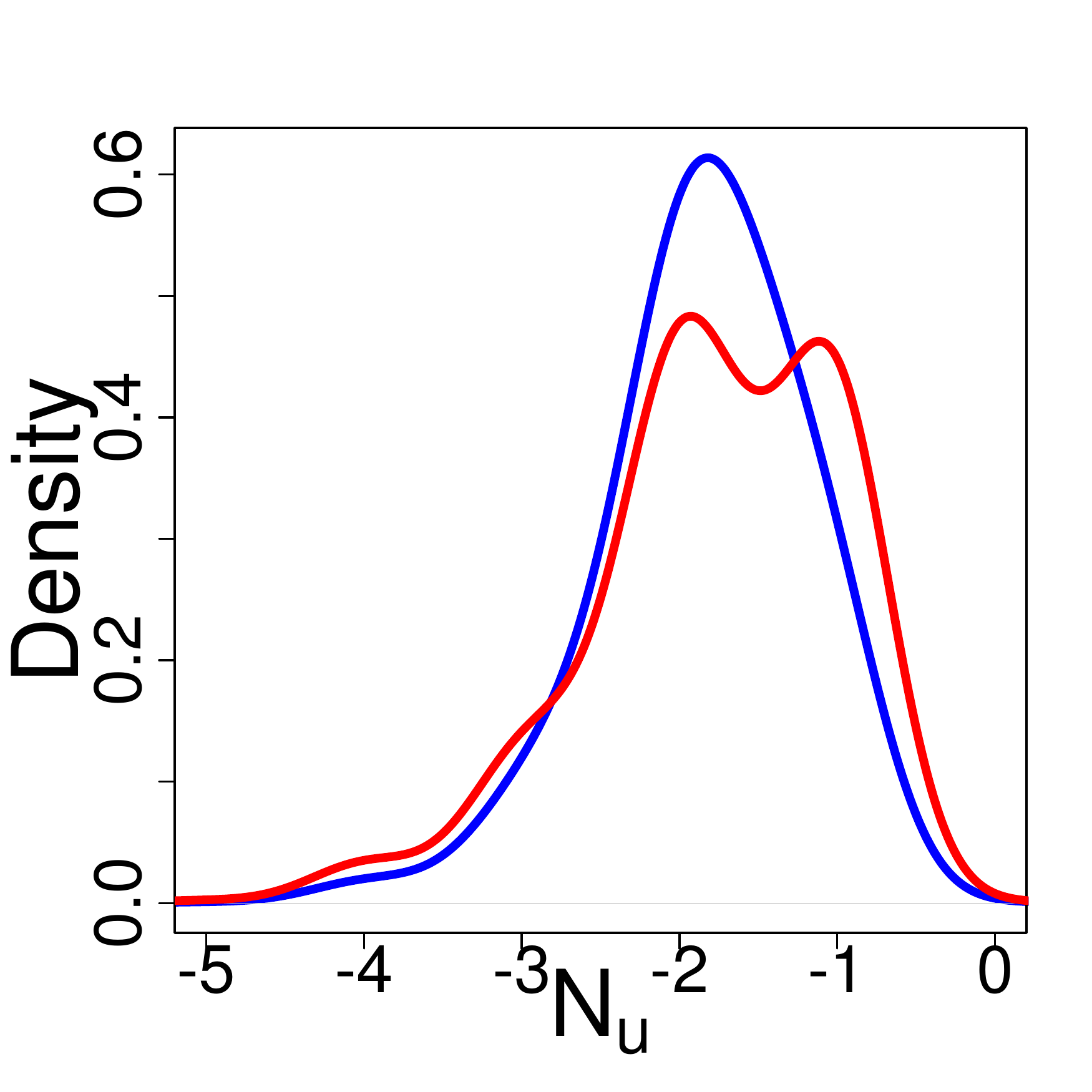}
\hfill  \includegraphics[width=0.5\textwidth]{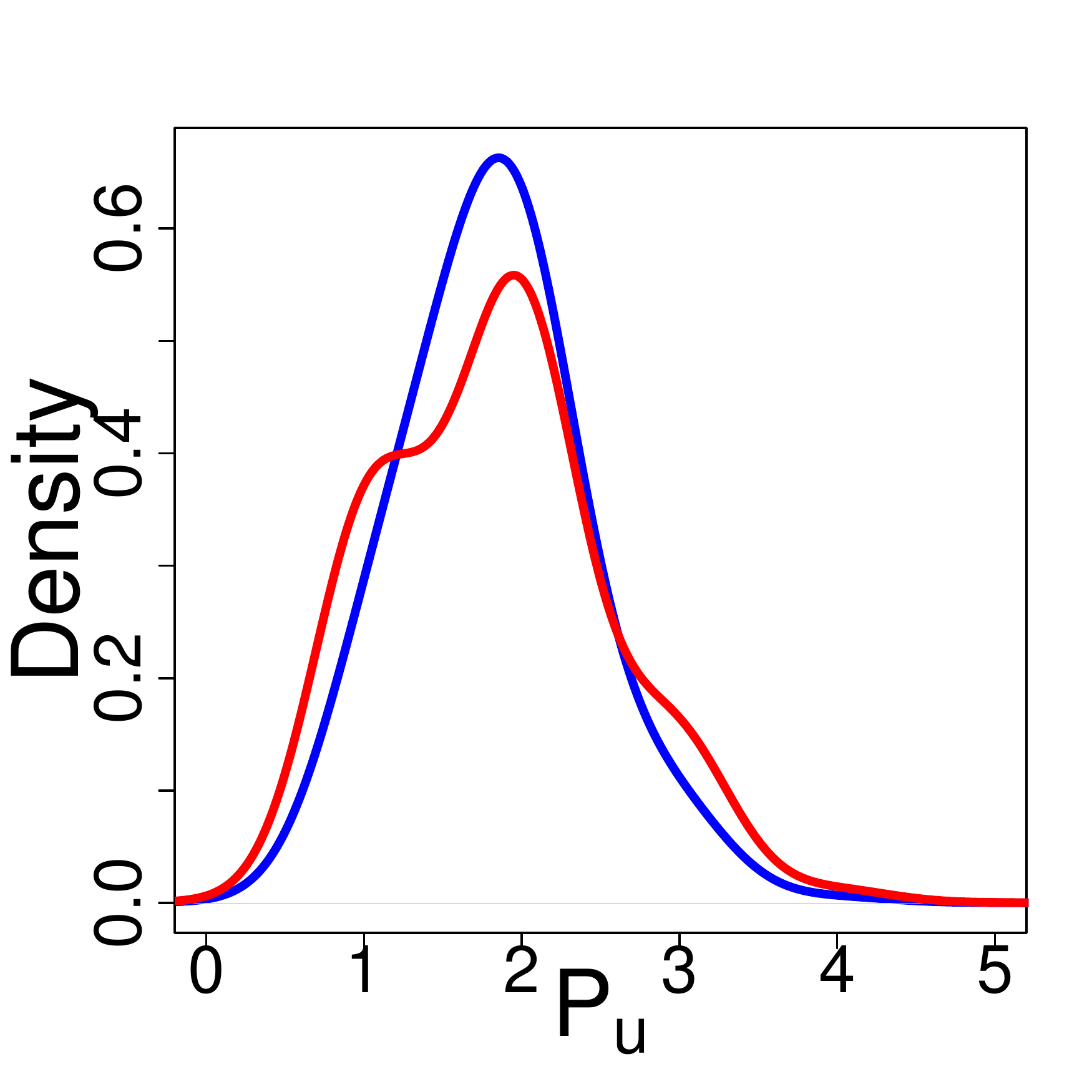}}
\caption{Conditional distributions of contributor emotions $P(N_u|I)$
  and $P(P_u|I)$) in the bug tracker (top) and the mailing list
  (bottom) for $I=\text{ACT}$ (blue) and $I=\text{INA}$
  (red). Distributions were smoothed through a Gaussian kernel of
  width $0.35$. 
  \label{fig:bgEmo}}
\end{figure}
 
We calculate the conditional distributions of emotions given the
label of an interval, $P(N_u|I)$ and $P(P_u|I)$, which we show in
Figure \ref{fig:bgEmo} for both datasets. An initial inspection shows
the differences between the expression of emotions when a contributor
is going to become inactive and when not. For both datasets, the
distribution of emotional expression followed by an interval labeled
as \emph{INA} has larger variance than when followed by intervals
labeled as \emph{ACT}, showing signs of bimodality. Wilcoxon tests
reveal that the conditional distributions of both emotions in the bug
tracker are significantly different ($p<1e-15$).
In the mailing list, this is the case only for $N_u$ ($p<1e-15$), while the null
hypothesis could not be rejected ($p=0.21$) for $P_u$. This
highlights the role of negative expression among developers, which
differs more when one is going to become inactive, in comparison with
active periods. Nevertheless, for the case of the mailing list, the failure to reject the null
hypothesis for $P_u$ does not imply that it is not informative, as we
show below. 

\subsection{Activity tendencies}

\begin{figure}[!ht] 
\centering
\centerline{ 
  \includegraphics[width=0.5\textwidth]{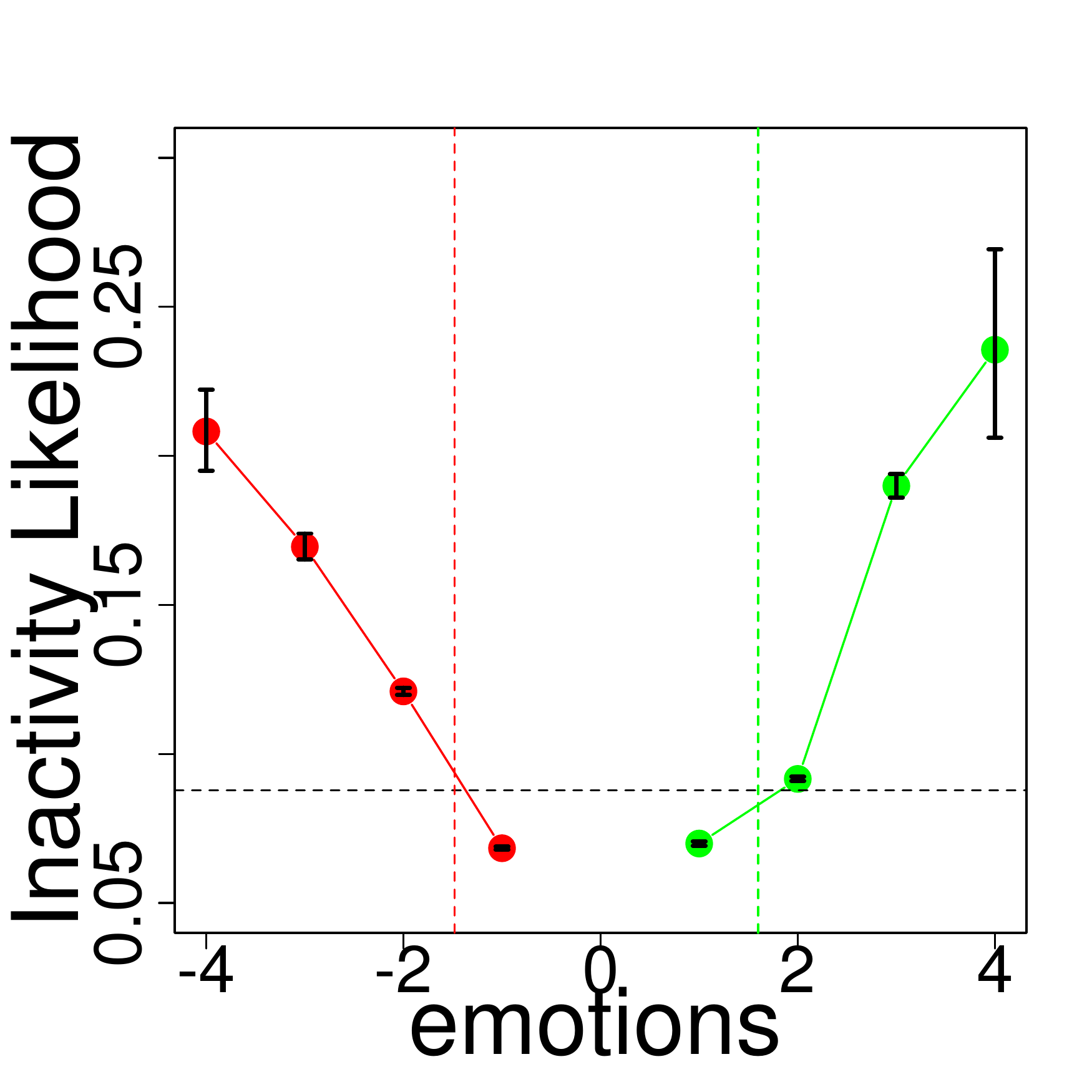}
\hfill  \includegraphics[width=0.5\textwidth]{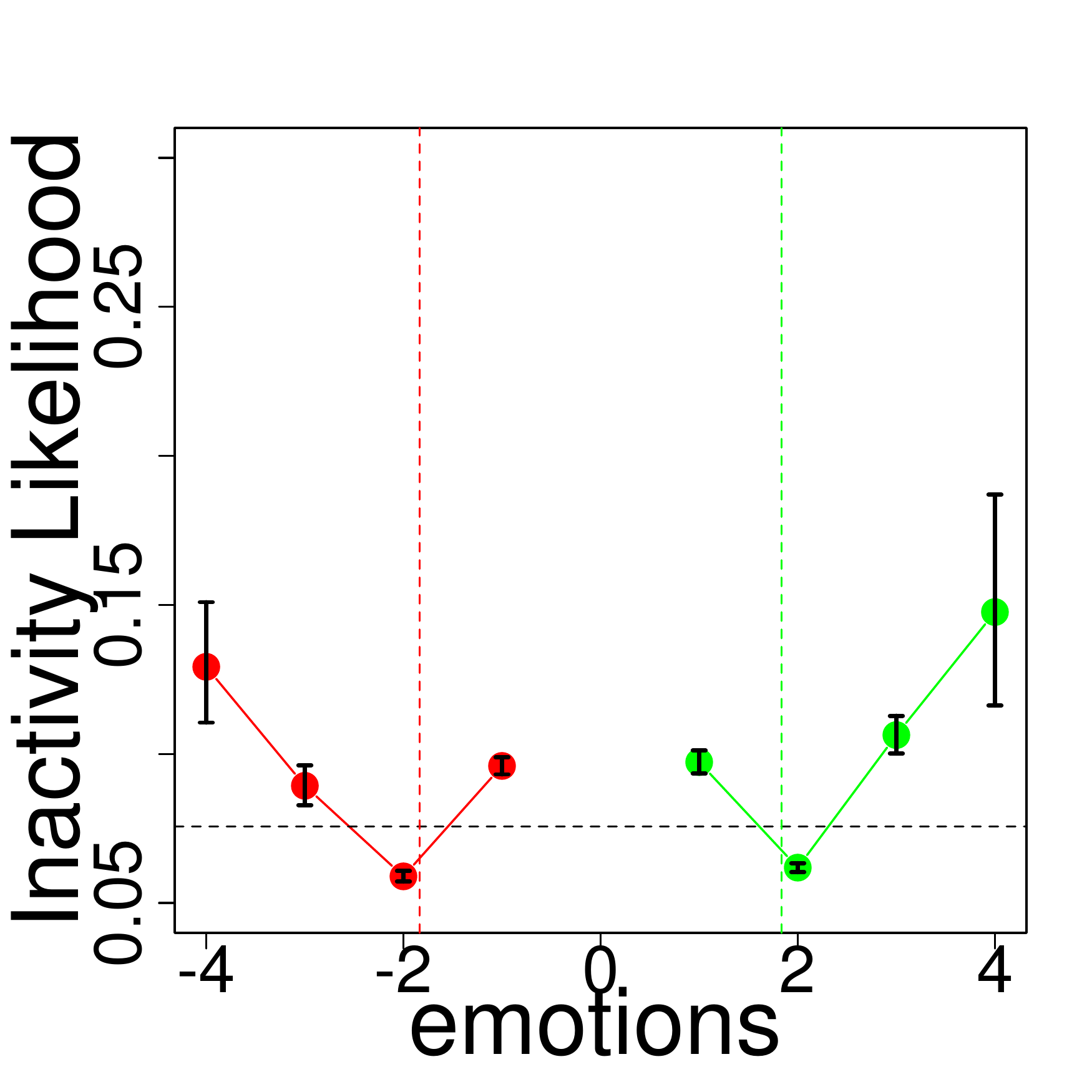}}
  \caption{ Likelihood of an interval to be labeled as \emph{INA} given contributor emotions, $P(I=\text{INA} | N_u)$ (red) and $P(I=\text{INA} | P_u)$ (green), for the bug tracker (left) and the mailing list (right). 
Error bars show confidence intervals, the horizontal dashed lines indicate $P(I=\text{INA})$ and vertical bars the means of $N_u$ and $P_u$ in each dataset.\label{fig:likelihood}}
\end{figure}

A notable difference in the distributions of Figure \ref{fig:bgEmo} is the range where $P(N_u|I=\text{INA})>P(N_u|I=\text{ACT})$ and $P(P_u|I=\text{INA})>P(P_u|I=\text{ACT})$. 
For the case of the bug tracker, this condition is present only when $N_u$ and $P_u$ are above a certain value (i.e. in terms of absolute valence), while for the case of the mailing list, this is also true when $N_u$ and $P_u$ have a sufficiently low absolute value. 
This indicates that strong emotions in the bug tracker, and deviations from the mean emotions in the mailing list inform about the likelihood of a contributor becoming inactive.
To measure these effects, we compute the posterior distribution of a contributor becoming inactive at given time, considering his emotional expression in the last five days as

\begin{equation}
P(I=\text{INA} | N_u) = \frac{ P(N_u | I=\text{INA}) \cdot P(I=\text{INA})}{P(N_u)}
\label{eq:bayes}
\end{equation}
   
and its equivalent for $P_u$. We bin $P_u$ and $N_u$ in five bins, using the ranges $[1,5]$ and $[-1,-5]$ respectively, computing confidence intervals for the posterior distribution. 
Figure \ref{fig:likelihood} shows the posterior likelihood of becoming inactive for the first four bins, as the fifth one was not giving significant values due to the low probability of having $|P_u|>4$ and $|N_u|>4$. 
Our observation of the difference of the influence of emotions in both communication channels becomes clear: the likelihood of becoming inactive increases with $P_u$ and $N_u$ in the case of the bug tracker, while it grows with the distance to the mean for the case of the mailing list.

The distinctive v-shape of the likelihood for the emotions in the mailing list (Figure \ref{fig:likelihood} right) implies that lack of emotions can also serve as an indicator for contributors becoming inactive, but only when shared with others through the mailing list. 
It is remarkable that both $N_u$ and $P_u$ are informative to discriminate periods of inactivity in both datasets, suggesting that the decision to become inactive is more related to emotional intensity in general, rather than to positive or negative emotions alone. 
This is in line with the psychological theory which states that certain levels of arousal, or emotion intensity, are motivators for activity \cite{Schweitzer2010}.
 
\subsection{Real-time prediction}  
We apply the Bayesian analysis explained above to predict when contributors are going to start periods of inactivity, solely based on the emotional content of their messages.  Given the results shown in Figure \ref{fig:likelihood}, we apply two different models:
\begin{enumerate} 
\item bug tracker: if $|N_u| > \Theta_1$  or $|P_u| > \Theta_1$, then the next interval is predicted to be \emph{INA}, and \emph{ACT} otherwise.
\item mailing list: if $|N_u - \bar{N_u}| > \Theta_2$ or $|P_u - \bar{P_u}| > \Theta_2$, then the next interval is predicted to be \emph{INA}, and \emph{ACT} otherwise, where $\bar{N_u}$ and $\bar{P_u}$ are the average values of emotions expressed by this contributor.
\end{enumerate}

We apply the above predictors with $\Theta_1=1.9$ and $\Theta_2=0.8$ to each point in our datasets, and compute values of \emph{Precision} and \emph{Recall} \cite{Hooimeijer2007} over $20$ bootstrapped samples, to ensure the robustness of our predictor. 
Table \ref{tab:Precision} reports the means and standard deviations of \emph{Precision} and \emph{Recall} for each class and dataset. 
Both predictors have \emph{Precision} significantly higher than the prior probabilities in the respective classes. 
In particular, the \emph{Precision} of the minority class, \emph{INA}, is sufficiently above the prior probability $P(I=\text{INA})$, showing that our method produce meaningful results when using contributor emotions to predict when these are at risk of becoming inactive. 
In addition, the values of \emph{Recall} for both classes are well above $0.6$, correctly classifying most of the existing instances.

\begin{table}[htpb]  
\centering 
\caption{Results of the prediction of contributors becoming inactive
  or remaining active in both datasets. Standard deviations of Precision
  and Recall values are calculated over $20$ bootstrapped samples of the
  datasets. \label{tab:Precision}} 
\begin{tabular}{c|c|c|c}
     Dataset & measure  & \emph{Active} & \emph{Inactive} \\
     \hline

     \rowcolor{Gray}      &   Prior probability   &  $0.912$    &  $0.088$    \\
                                         &   Precision           &
                                         $0.947 \pm 0.019$   &
                                     $0.156 \pm 0.015$ \\
     \rowcolor{Gray}     \multirow{-3}{*}{ \textsc{Gentoo} \textsc{Bugzilla}}                   &   Recall              &  $0.673      \pm 0.087 $ &  $0.625 \pm 0.013 $\\
\hline
\hline
 
     \rowcolor{Gray}  &   Prior probability           &  $0.924$    &  $0.075$    \\
                    &   Precision                   &  $0.954 \pm 0.019$   & $0.129 \pm 0.025$ \\
     \rowcolor{Gray}         \multirow{-3}{*}{ \textsc{gentoo-dev}}   &  Recall                      &  $0.655 \pm 0.073$ &  $0.635 \pm 0.022$\\
\end{tabular}
\end{table}  

\section{Conclusions} 
     
Based on a case study of the \textsc{Gentoo} project, we analyzed the relation between emotions and activity of its contributors.  We gathered two disjoint datasets of communication
within the community: (i) the bug reports stored in
its bug tracking system \textsc{bugzilla}, and (ii) the messages posted
in the developer's mailing list.  We provided a sentiment analysis of the messages written by
all the contributors and related the emotional expressions to the activity patterns. 

The first part of our case study investigated the emotional components related to the leave of a central contributor, named \emph{Alice}. We show that her email discussions with other
contributors were more negative than the
rest, and that her departure was followed by higher stages of
negativity in the community during its reorganization.

We extended this analysis to contributors in general, both in the bug
tracker and the mailing list.  To detect inactivity, we applied
current state-of-the art theories on human correlated behavior,
finding a mode of interevent times that indicates stages of
contributor inactivity.  This allowed us to statistically analyze the
relation between a contributor's emotional expressions and its individual
intervals of inactivity.  We  reveal preconditions of emotional
expressions that indicate when contributors feel demotivated to further
contribute to the project.  Based on this, we are able to estimate when
a contributor becomes inactive, based on emotions expressed on his
last messages. With this, we  provide a tractable approach that can be applied by community managers to monitor emotional interaction within the community, and to
foster timely reaction against undesirable turnover events of contributors.

Our contributions do not only focus on predictive results, but provide additional insights into the phenomenon at hand, in
particular into fundamental relations between emotions and activity (and implicitly into motivation). We find that it is the
emotional \emph{intensity} which defines activity, rather than its \emph{polarity}
in terms of positive or negative emotions.  Thus, in this work, we
took a step forward by providing a methodology based on
\emph{sentiment} analysis, which sheds new light on \textsc{Gentoo}'s
case study.  This unveils a wide horizon of new quantitative
approaches to the analysis of social dynamics within online
communities, extending previous approaches to online emotional
interaction \cite{Garas2012}, and social resilience
\cite{Garcia2013a}.

\section*{Acknowledgment}
The research leading to these results has received funding from the
European Community's Seventh Framework Programme FP7-ICT-2008-3 under
grant agreement no 231323 (CYBEREMOTIONS).

\bibliographystyle{IEEEtran}
  
\bibliography{emotionoss}   

\begin{thebibliography}{10}
\providecommand{\url}[1]{#1}
\csname url@samestyle\endcsname
\providecommand{\newblock}{\relax}
\providecommand{\bibinfo}[2]{#2}
\providecommand{\BIBentrySTDinterwordspacing}{\spaceskip=0pt\relax}
\providecommand{\BIBentryALTinterwordstretchfactor}{4}
\providecommand{\BIBentryALTinterwordspacing}{\spaceskip=\fontdimen2\font plus
\BIBentryALTinterwordstretchfactor\fontdimen3\font minus
  \fontdimen4\font\relax}
\providecommand{\BIBforeignlanguage}[2]{{%
\expandafter\ifx\csname l@#1\endcsname\relax
\typeout{** WARNING: IEEEtran.bst: No hyphenation pattern has been}%
\typeout{** loaded for the language `#1'. Using the pattern for}%
\typeout{** the default language instead.}%
\else
\language=\csname l@#1\endcsname
\fi
#2}}
\providecommand{\BIBdecl}{\relax}
\BIBdecl

\bibitem{Andreoni1990}
J.~Andreoni, ``{Impure Altruism and Donations to Public Goods: A Theory of
  Warm-Glow Giving},'' \emph{The Economic Journal}, vol. 100, no. 401, p. 464,
  1990.

\bibitem{Hardin1968}
G.~Hardin, ``{The Tragedy of the Commons},'' \emph{Science}, vol. 162, no.
  5364, pp. 1243--8, 1968.

\bibitem{crowston2006}
K.~Crowston and K.~Wei, ``{Core and periphery in Free/Libre and Open Source
  software team communications},'' \emph{System Sciences, 2006.}, vol.~00,
  no.~C, pp. 1--7, 2006.

\bibitem{Ehrlich2007}
K.~Ehrlich, G.~Valetto, and M.~Helander, ``{Seeing inside: Using social network
  analysis to understand patterns of collaboration and coordination in global
  software teams},'' \emph{ICGSE 2007}, no. Icgse, pp. 297--298, Aug. 2007.

\bibitem{He2012}
P.~He, B.~Li, and Y.~Huang, ``{Applying Centrality Measures to the Behavior
  Analysis of Developers in Open Source Software Community},'' \emph{Second
  IEEE International Conference on Cloud and Green Computing}, pp. 418--423,
  2012.

\bibitem{Zanetti2013a}
M.~S. Zanetti, I.~Scholtes, C.~J. Tessone, and F.~Schweitzer, ``Categorizing
  bugs with social networks: A case study on four open source software
  communities,'' in \emph{Proceedings of the ICSE '13}, 2013, pp. 1032--1041.

\bibitem{thelwall2012sentiment}
M.~Thelwall, K.~Buckley, and G.~Paltoglou, ``Sentiment strength detection for
  the social web,'' \emph{Journal of the American Society for Information
  Science and Technology}, vol.~63, no.~1, pp. 163--173, 2012.

\bibitem{Crowston2012}
K.~Crowston, K.~Wei, J.~Howison, and A.~Wiggins, ``Free/libre open-source
  software development: What we know and what we do not know,'' \emph{ACM
  Computing Surveys}, vol.~44, no.~2, pp. 1--35, 2012.

\bibitem{serrano2012co}
M.~S. Zanetti, ``The co-evolution of socio-technical structures in sustainable
  software development: Lessons from the open source software communities,'' in
  \emph{Proceedings of the 34th ICSE}.\hskip 1em plus 0.5em minus 0.4em\relax
  IEEE Press, 2012, pp. 1587--1590.

\bibitem{valetto2007using}
G.~Valetto, M.~Helander, K.~Ehrlich, S.~Chulani, M.~Wegman, and C.~Williams,
  ``Using software repositories to investigate socio-technical congruence in
  development projects,'' in \emph{MSR '07}.\hskip 1em plus 0.5em minus
  0.4em\relax IEEE, 2007, pp. 25--25.

\bibitem{Mockus2002tse}
A.~Mockus, R.~T. Fielding, and J.~D. Herbsleb, ``{Two case studies of open
  source software development: Apache and Mozilla},'' \emph{ACM Transactions on
  Software Engineering and Methodology}, vol.~11, no.~3, pp. 309--346, 2002.

\bibitem{lerner2002}
J.~Lerner and J.~Tirole, ``Some simple economics of open source,''
  \emph{Journal of Industrial Economics}, pp. 197--234, 2002.

\bibitem{gacek2004}
C.~Gacek and B.~Arief, ``The many meanings of open source,'' \emph{IEEE
  Software}, vol.~21, pp. 34--40, 2004.

\bibitem{krogh2006}
G.~von Krogh and E.~von Hippel, ``The promise of research on open source
  software,'' \emph{Management Science}, vol.~52, no.~7, p. 975, 2006.

\bibitem{cataldo2008}
M.~Cataldo, J.~D. Herbsleb, and K.~M. Carley, ``Socio-technical congruence: a
  framework for assessing the impact of technical and work dependencies on
  software development productivity,'' in \emph{ESEM}, 2008.

\bibitem{Kappas2013}
A.~Kappas, ``{Social regulation of emotion: messy layers.}'' \emph{Frontiers in
  psychology}, vol.~4, no. February, p.~51, 2013.

\bibitem{Chmiel2011}
A.~Chmiel, J.~Sienkiewicz, M.~Thelwall, G.~Paltoglou, K.~Buckley, A.~Kappas,
  and J.~A. Hołyst, ``{Collective Emotions Online and Their Influence on
  Community Life},'' \emph{PLoS ONE}, vol.~6, no.~7, p. e22207, 2011.

\bibitem{Garas2012}
A.~Garas, D.~Garcia, M.~Skowron, and F.~Schweitzer, ``{Emotional persistence in
  online chatting communities},'' \emph{Scientific Reports}, vol.~2, p. 402,
  2012.

\bibitem{Garcia2011}
D.~Garcia, A.~Garas, and F.~Schweitzer, ``{Positive words carry less
  information than negative words},'' \emph{EPJ Data Science}, vol.~1, no.~1,
  p.~3, 2012.

\bibitem{Garcia2012b}
D.~Garcia and F.~Schweitzer, ``{Modeling online collective emotions},'' in
  \emph{Proceedings of the 2012 workshop on Data-driven user behavioral
  modelling and mining from social media - DUBMMSM '12}, 2012, p.~37.

\bibitem{Bollen2010}
J.~Bollen, H.~Mao, and X.-j. Zeng, ``{Twitter mood predicts the stock
  market},'' \emph{Journal of Computational Science}, vol.~2, pp. 1--8, 2011.

\bibitem{Deng2011}
S.~Deng, T.~Mitsubuchi, K.~Shioda, T.~Shimada, and A.~Sakurai, ``{Combining
  Technical Analysis with Sentiment Analysis for Stock Price Prediction},''
  \emph{2011 IEEE Ninth International Conference on Dependable, Autonomic and
  Secure Computing}, pp. 800--807, 2011.

\bibitem{Garcia2011e}
D.~Garcia and F.~Schweitzer, ``{Emotions in Product Reviews – Empirics and
  Models},'' \emph{Proceedings of 2011 IEEE International Conference on Social
  Computing, SocialCom}, pp. 483--488, 2011.

\bibitem{Pfitzner2012}
R.~Pfitzner and A.~Garas, ``{Emotional divergence influences information
  spreading in Twitter},'' in \emph{AAAI ICWSM 2012}, 2012, pp. 2--5.

\bibitem{Paltoglou2010}
G.~Paltoglou, S.~Gobron, M.~Skowron, M.~Thelwall, and D.~Thalmann, ``{Sentiment
  analysis of informal textual communication in cyberspace},'' in \emph{In
  Proc. Engage 2010}, 2010, pp. 13--25.

\bibitem{Zhang2012}
L.~Zhang, Y.~Jia, B.~Zhou, and Y.~Han, ``{Microblogging Sentiment Analysis
  Using Emotional Vector},'' \emph{2012 IEEE Second International Conference on
  Cloud and Green Computing}, pp. 430--433, 2012.

\bibitem{Thelwall2013}
M.~Thelwall, K.~Buckley, G.~Paltoglou, M.~Skowron, D.~Garcia, S.~Gobron,
  J.~Ahn, A.~Kappas, D.~Kuster, and A.~Janusz, ``{Damping Sentiment Analysis in
  Online Communication : Discussions , Monologs and Dialogs},'' in
  \emph{Proceedings of the 25th International Conference on Computational
  Linguistics (COLING)}, 2013, pp. 1--12.

\bibitem{Garcia2012a}
D.~Garcia, F.~Mendez, U.~Serd\"{u}lt, and F.~Schweitzer, ``{Political
  polarization and popularity in online participatory media},'' in
  \emph{Proceedings of the workshop on Politics, elections and data}, 2012, pp.
  3--10.

\bibitem{Wu2011}
Y.~Wu, J.~Wong, Y.~Deng, and K.~Chang, ``{An Exploration of Social Media in
  Public Opinion Convergence: Elaboration Likelihood and Semantic Networks on
  Political Events},'' \emph{2011 IEEE Ninth International Conference on
  Dependable, Autonomic and Secure Computing}, pp. 903--910, 2011.

\bibitem{Kucuktunc2012}
O.~Kucuktunc, B.~B. Cambazoglu, I.~Weber, and H.~Ferhatosmanoglu, ``{A
  large-scale sentiment analysis for Yahoo! answers},'' in \emph{Proceedings of
  the fifth ACM international conference on Web search and data mining - WSDM
  '12}, New York, New York, USA, 2012, p. 633.

\bibitem{Backstrom2006}
L.~Backstrom, D.~Huttenlocher, J.~Kleinberg, and X.~Lan, ``{Group formation in
  large social networks},'' in \emph{Proceedings of the 12th international
  conference on Knowledge discovery and data mining - KDD '06}, 2006, p.~44.

\bibitem{Zheleva2009}
E.~Zheleva, H.~Sharara, and L.~Getoor, ``{Co-evolution of social and
  affiliation networks},'' in \emph{Proceedings of the 15th ACM SIGKDD
  international conference on Knowledge discovery and data mining - KDD
  '09}.\hskip 1em plus 0.5em minus 0.4em\relax ACM Press, 2009, p. 1007.

\bibitem{Walter2009}
F.~E. Walter, S.~Battiston, and F.~Schweitzer, ``{Personalised and dynamic
  trust in social networks},'' in \emph{Proceedings of the third ACM conference
  on Recommender systems - RecSys '09}, 2009, pp. 197--204.

\bibitem{Gonzalez-Bailon2011}
S.~Gonz\'{a}lez-Bail\'{o}n, J.~Borge-Holthoefer, A.~Rivero, and Y.~Moreno,
  ``{The dynamics of protest recruitment through an online network.}''
  \emph{Scientific reports}, vol.~1, p. 197, 2011.

\bibitem{Dror2012}
G.~Dror, D.~Pelleg, O.~Rokhlenko, and I.~Szpektor, ``{Churn prediction in new
  users of Yahoo! answers},'' in \emph{Proceedings of the 21st international
  conference companion on World Wide Web - WWW '12 Companion}, New York, New
  York, USA, 2012, p. 829.

\bibitem{Wu2013}
S.~Wu, A.~{Das Sarma}, A.~Fabrikant, S.~Lattanzi, and A.~Tomkins, ``{Arrival
  and departure dynamics in social networks},'' in \emph{Proceedings of the
  sixth ACM international conference on Web search and data mining - WSDM '13},
  New York, New York, USA, 2013, p. 233.

\bibitem{Herrera2007}
O.~Herrera and T.~Znati, ``{Modeling Churn in P2P Networks},'' in \emph{40th
  IEEE Annual Simulation Symposium (ANSS'07)}, 2007, pp. 33--40.

\bibitem{Karnstedt2010}
M.~Karnstedt, T.~Hennessy, J.~Chan, and C.~Hayes, ``{Churn in Social Networks:
  A Discussion Boards Case Study},'' in \emph{2010 IEEE Second International
  Conference on Social Computing}, 2010, pp. 233--240.

\bibitem{Kawale2009}
J.~Kawale, A.~Pal, and J.~Srivastava, ``{Churn Prediction in MMORPGs: A Social
  Influence Based Approach},'' in \emph{IEEE International Conference on
  Computational Science and Engineering}, 2009, pp. 423--428.

\bibitem{hall1961psychology}
J.~F. Hall, \emph{Psychology of motivation.}\hskip 1em plus 0.5em minus
  0.4em\relax Lippincott, 1961.

\bibitem{Garcia2013a}
\BIBentryALTinterwordspacing
D.~Garcia, P.~Mavrodiev, and F.~Schweitzer, ``{Social Resilience in Online
  Communities: The Autopsy of Friendster},'' 2013, under review. [Online].
  Available: \url{http://arxiv.org/abs/1302.6109}
\BIBentrySTDinterwordspacing

\bibitem{Cvijikj2011}
I.~P. Cvijikj and F.~Michahelles, ``{Monitoring Trends on Facebook},''
  \emph{2011 IEEE Ninth International Conference on Dependable, Autonomic and
  Secure Computing}, pp. 895--902, 2011.

\bibitem{Michalski2012}
R.~Michalski, J.~Jankowski, and P.~Kazienko, ``{Negative Effects of
  Incentivised Viral Campaigns for Activity in Social Networks},'' \emph{2012
  IEEE Second International Conference on Cloud and Green Computing}, pp.
  391--398, 2012.

\bibitem{serrano2005}
N.~Serrano and I.~Ciordia, ``Bugzilla, itracker, and other bug trackers,''
  \emph{Software, IEEE}, vol.~22, no.~2, pp. 11--13, 2005.

\bibitem{Zanetti2013b}
M.~S. Zanetti, I.~Scholtes, C.~J. Tessone, and F.~Schweitzer, ``The rise and
  fall of a central contributor: Dynamics of social organization and
  performance in the gentoo community,'' in \emph{Proceedings of the CHASE '13
  ICSE Workshop}, 2013, pp. 49--56.

\bibitem{Zanetti2012}
M.~S. Zanetti, E.~Sarigol, I.~Scholtes, C.~J. Tessone, and F.~Schweitzer, ``A
  quantitative study of social organisation in open source software
  communities,'' in \emph{Proceedings of the ICCSW '12 - Imperial College
  Computing Student Workshop}, vol.~28.\hskip 1em plus 0.5em minus 0.4em\relax
  Schloss Dagstuhl, 2012, pp. 116--122.

\bibitem{mockus2012}
M.~Zhou and A.~Mockus, ``What make long term contributors: Willingness and
  opportunity in oss community,'' in \emph{ICSE}, 2012, pp. 518--528.

\bibitem{Schweitzer2010}
F.~Schweitzer and D.~Garcia, ``An agent-based model of collective emotions in
  online communities,'' \emph{The European Physical Journal B}, vol.~77, no.~4,
  pp. 533--545, 2010.

\bibitem{Barabasi2005}
A.-L. Barabasi, ``{The origin of bursts and heavy tails in human dynamics.}''
  \emph{Nature}, vol. 435, no. 7039, pp. 207--11, 2005.

\bibitem{Wu2010}
Y.~Wu, C.~Zhou, J.~Xiao, J.~Kurths, and H.~J. Schellnhuber, ``{Evidence for a
  bimodal distribution in human communication.}'' \emph{Proceedings of the
  National Academy of Sciences of the United States of America}, vol. 107,
  no.~44, pp. 18\,803--8, 2010.

\bibitem{Hooimeijer2007}
P.~Hooimeijer and W.~Weimer, ``Modeling bug report quality,'' in
  \emph{Proceedings of the 22nd IEEE/ACM international conference on Automated
  software engineering}, 2007, pp. 34--43.

\end{thebibliography}
  
\end{document}